\documentclass[aps,pra,twocolumn]{revtex4}%
\usepackage{graphics}
\usepackage{amsmath}
\usepackage{graphicx}
\usepackage{amsfonts}
\usepackage{amssymb}
\usepackage{float}
\usepackage{longtable}
\usepackage{epsfig}
\usepackage{latexsym}
\usepackage{theorem}
\usepackage{bbm}
\usepackage{bm}
\usepackage{psfrag}
\newtheorem{theorem}{Theorem}
\newtheorem{lemma}[theorem]{Lemma}

\newtheorem{remark}[theorem]{Remark}

\begin{document}
\title{Teleportation simulation of bosonic Gaussian channels:\\Strong and uniform convergence}
\author{Stefano Pirandola}
\affiliation{Computer Science and York Centre for Quantum Technologies, University of York,
York YO10 5GH, UK}
\author{Riccardo Laurenza}
\affiliation{Computer Science and York Centre for Quantum Technologies, University of York,
York YO10 5GH, UK}
\author{Samuel L. Braunstein}
\affiliation{Computer Science and York Centre for Quantum Technologies, University of York,
York YO10 5GH, UK}

\begin{abstract}
We consider the Braunstein-Kimble protocol for continuous variable
teleportation and its application for the simulation of bosonic channels. We
discuss the convergence properties of this protocol under various topologies
(strong, uniform, and bounded-uniform) clarifying some typical
misinterpretations in the literature. We then show that the teleportation
simulation of an arbitrary single-mode Gaussian channel is uniformly
convergent to the channel if and only if its noise matrix has full rank. The
various forms of convergence are then discussed within adaptive protocols,
where the simulation error must be propagated to the output of the protocol by
means of a \textquotedblleft peeling\textquotedblright\ argument, following
techniques from PLOB [arXiv:1510.08863]. Finally, as an application of the
peeling argument and the various topologies of convergence, we provide
complete rigorous proofs for recently-claimed strong converse bounds for
private communication over Gaussian channels.

\end{abstract}
\maketitle

\section{Introduction}

Quantum teleportation~\cite{Tele1,Tele2,telereview,oldrev,teleCV2} is a
fundamental operation in quantum information
theory~\cite{NielsenChuang,CVbook,RMP} and quantum Shannon
theory~\cite{HolevoBOOK,Hayashi}. It is a central tool for simulating quantum
channels with direct applications to quantum/private communications~\cite{TQC}
and quantum metrology~\cite{ReviewMETRO}. In a seminal paper, Bennett et
al.~\cite{ref1} showed how to simulate Pauli channels and reduce quantum
communication protocols into entanglement distillation. Similar ideas can be
found in a number of other
investigations~\cite{ref2,ref3,ref4,ref5,ref6,ref7,ref8,ref9,ref10,ref11,ref12,ref13,ref14,ref15}
(see Ref.~\cite[Sec.~IX]{TQC} for a detailed discussion of the literature on
channel simulation). More recently, in 2015,
Pirandola-Laurenza-Ottaviani-Banchi (PLOB)~\cite{PLOB} showed how to transform
these precursory ideas into a completely general formulation.

PLOB showed how to simulate an arbitrary quantum channel (in arbitrary
dimension) by means of local operations and classical communication (LOCC)
applied to the channel input and a suitable resource state. For instance, this
approach allowed one to deterministically simulate the amplitude damping
channel for the very first time. The LOCC\ simulation of a quantum channel is
then exploited in the technique of teleportation stretching~\cite{PLOB}, where
an arbitrary adaptive protocol (i.e., based on the use of feedback) is
simplified into a simpler block version, where no feedback is involved.

Teleportation stretching is a very flexible technique whose combination with
suitable entanglement measures\ (such as the relative entropy of
entanglement~\cite{REE1,REE2,REE3}) and other functionals (such as the quantum
Fisher information~\cite{qfi1,qfi2,qfi3,qfi4,qfi5}) has recently led to the
discovery of a number of results. For instance, PLOB established the two-way
assisted quantum/private capacities of various fundamental channels, such as
the lossy channel, the quantum-limited amplifier, dephasing and erasure
channels~\cite{PLOB}. In particular, the PLOB bound of $-\log(1-\tau)$ bits
per use of a lossy channel with transmissivity $\tau$ sets the ultimate limit
of point-to-point quantum communications or, equivalently, a fundamental
benchmark for quantum
repeaters~\cite{Briegel,Rep2,Rep3,Rep4,Rep5,Rep6,Rep7,Rep8,Rep9,Rep10,Rep12,Rep13,Rep13bis,Rep14,Rep15,Rep16,Rep17,Rep18,Rep19}%
. In the setting of quantum metrology, Ref.~\cite{PirCo} used teleportation
stretching to show that parameter estimation with teleportation-covariant
channels cannot beat the standard quantum limit, establishing the adaptive
limits achievable in many scenarios. Other results were established for
quantum networks~\cite{netPAPER}, such as a quantum version of the max
flow/min cut theorem. See also Refs.~\cite{next1,next2,next3,next4} for other studies.

It is clear that continuous variable (CV) quantum
teleportation~\cite{telereview}, also known as the Braunstein-Kimble (BK)
protocol~\cite{Tele2}, is central in many of the previous results and in
several other important applications. The BK protocol is a tool for optical
quantum communications, from realistic implementations of quantum key
distribution, e.g., via swapping in untrusted
relays~\cite{MDI1,CVMDIQKD,CVMDIQKD2,CVMDIQKD3} to more ambitious goals such
as the design of a future quantum Internet~\cite{HybridINTERNET,Kimble2008}.
That being said, the BK\ protocol is still the subject of misunderstandings by
some authors. Typical misuses arise from confusing the different forms of
convergence that can be associated with this protocol, an error which is
connected with a specific order of the limits to be carefully considered when
teleportation is performed within an infinite-dimensional Hilbert space.

In this work, we discuss and clarify the convergence properties of the
BK\ protocol and its consequences for the simulation of bosonic channels. As a
specific case, we investigate the simulation of single-mode bosonic Gaussian
channels, which can be fully classified in different canonical
forms~\cite{HolevoCanonical,Caruso,HolevoVittorio} up to input/output Gaussian
unitaries. We show that the teleportation simulation of a single-mode Gaussian
channel uniformly converges to the channel as long as its noise matrix has
full rank. This matrix is generally connected with the covariance matrix of
the Gaussian state describing the environment in a single-mode symplectic
dilation of the quantum channel.

Assuming various topologies of convergence (strong, uniform, and
bounded-uniform), we then study the teleportation simulation of bosonic
channels in adaptive protocols. Here we discuss the crucial role of a peeling
argument that connects the channel simulation error, associated with the
single channel transmissions, to the overall simulation error accumulated on
the final quantum state at the output of the protocol. This argument is needed
in order to rigorously prove \textit{strong converse} upper bounds for two-way
assisted private capacities. As a direct application of our analysis, we then
provide various complete proofs for the strong converse bounds claimed in
Wilde-Tomamichel-Berta (WTB)~\cite{WildeFollowup}. In particular, we show how
the bounds claimed in WTB can be rigorously proven for adaptive protocols, and
how their illness (divergence to infinity) is fixed by a correct use of the
BK\ teleportation protocol. In this regard, our study extends the one already
given in Ref.~\cite{TQC} to also include the topologies of strong and uniform convergence.

The paper is organized as follows. In Sec.~II, we provide some preliminary
notions on bosonic systems, Gaussian states, and Gaussian channels, including
the classification in canonical
forms~\cite{HolevoCanonical,Caruso,HolevoVittorio}, as revisited in terms of
matrix ranks in Ref.~\cite{RMP}. In Sec.~III, we discuss the convergence
properties of the BK protocol for CV\ teleportation, also discussing the
interplay between the different limits associated with this protocol. In
Sec.~IV, we consider the teleportation simulation of bosonic channels under
the topologies of strong and bounded-uniform convergence. In Sec.~V, we
present the main result of our work, which is the necessary and sufficient
condition for the uniform convergence of the teleportation simulation of a
Gaussian channel. In Sec.~V, we present the peeling argument for adaptive
protocols, considering the various forms of convergence. Next, in Sec.~VI, we
present implications for quantum/private communications, showing the rigorous
proofs of the claims presented in WTB. Finally, Sec.~VII\ is for conclusions.

\section{Preliminaries}

\subsection{Bosonic systems and Gaussian states\label{SecII}}

CV systems have an infinite-dimensional Hilbert space $\mathcal{H}$. The most
important example of CV systems is given by the bosonic modes of the radiation
field. In general, a bosonic system of $n$ modes is described by a tensor
product Hilbert space $\mathcal{H}^{\otimes n}$ and a vector of quadrature
operators $\mathbf{\hat{x}}^{T}:=(\hat{q}_{1},\hat{p}_{1},\ldots,\hat{q}%
_{n},\hat{p}_{n})$ satisfying the commutation relations
\begin{equation}
\lbrack\hat{x}_{l},\hat{x}_{m}]=2i\Omega_{lm}~~(1\leq l,m\leq2n)~, \label{CCR}%
\end{equation}
where $\mathbf{\Omega}$ is the symplectic form%
\begin{equation}
\mathbf{\Omega}:=\bigoplus\limits_{k=1}^{n}\boldsymbol{\omega}%
~,~\boldsymbol{\omega}:=\left(
\begin{array}
[c]{cc}%
0 & 1\\
-1 & 0
\end{array}
\right)  . \label{Symplectic_Form}%
\end{equation}

An arbitrary bosonic state is characterized by a density operator $\rho
\in\mathcal{D}(\mathcal{H}^{\otimes n})$ or, equivalently, by its Wigner
representation. Introducing the Weyl operator \cite{Weyl}
\begin{equation}
\hat{D}(\boldsymbol{\xi}):=\exp(i\mathbf{\hat{x}}^{T}\boldsymbol{\xi
}),~~\boldsymbol{\xi}\in\mathbb{R}^{2n}, \label{Weyl operator}%
\end{equation}
an arbitrary $\rho$ is equivalent to a characteristic function
\begin{equation}
\chi(\boldsymbol{\xi})=\mathrm{Tr}\left[  \rho\hat{D}(\boldsymbol{\xi
})\right]  ~, \label{CH_function}%
\end{equation}
or to a Wigner function%
\begin{equation}
W(\mathbf{x})=\int\limits_{\mathbb{R}^{2n}}\frac{d^{2n}\boldsymbol{\xi}}%
{(2\pi)^{2n}}~\exp\left(  -i\mathbf{x}^{T}\boldsymbol{\xi}\right)
\chi(\boldsymbol{\xi})~, \label{Wig_function}%
\end{equation}
where the continuous variables $\mathbf{x}^{T}:=(q_{1},p_{1},\ldots
,q_{n},p_{n})$ span the real symplectic space $\mathcal{K}:=(\mathbb{R}%
^{2n},\mathbf{\Omega})$ which is called the \emph{phase space}.

The most relevant quantities that characterize the Wigner representations are
the statistical moments. In particular, the first moment is the mean value%
\begin{equation}
\mathbf{\bar{x}}:=\left\langle \mathbf{\hat{x}}\right\rangle =\mathrm{Tr}%
(\mathbf{\hat{x}}\rho)~, \label{Displacement_1mom}%
\end{equation}
and the second moment is the covariance matrix (CM) $\mathbf{V}$, whose
arbitrary element is defined by%
\begin{equation}
V_{lm}:=\tfrac{1}{2}\left\langle \left\{  \Delta\hat{x}_{l},\Delta\hat{x}%
_{m}\right\}  \right\rangle ~,
\end{equation}
where $\Delta\hat{x}_{l}:=\hat{x}_{l}-\left\langle \hat{x}_{l}\right\rangle $
and $\{,\}$ is the anti-commutator. The CM is a $2n\times2n$, real symmetric
matrix which must satisfy the uncertainty principle%
\begin{equation}
\mathbf{V}+i\mathbf{\Omega}\geq0~, \label{BONA_FIDE}%
\end{equation}
coming directly from Eq.~(\ref{CCR}). For a particular class of states, the
first two moments are sufficient for a complete characterization. These are
the Gaussian states which, by definition, are those bosonic states whose
Wigner representation ($\chi$ or $W$) is Gaussian, i.e.,%
\begin{align}
\chi(\boldsymbol{\xi})  &  =\exp\left[  -\frac{1}{2}\boldsymbol{\xi}%
^{T}\mathbf{V}\boldsymbol{\xi}+i\mathbf{\bar{x}}^{T}\boldsymbol{\xi}\right]
~,\\
W(\mathbf{x})  &  =\frac{\exp\left[  -\frac{1}{2}(\mathbf{x}-\mathbf{\bar{x}%
})^{T}\mathbf{V}^{-1}(\mathbf{x}-\mathbf{\bar{x}})\right]  }{(2\pi)^{n}%
\sqrt{\det\mathbf{V}}}~.
\end{align}

It is also very important to identify the quantum operations that preserve the
Gaussian character of such quantum states. In the Heisenberg picture, Gaussian
unitaries correspond to canonical linear unitary Bogoliubov transformations,
i.e., affine real maps of the quadratures%
\begin{equation}
(\mathbf{S},\mathbf{d}):\mathbf{\hat{x}}\rightarrow\mathbf{S\hat{x}+d}~,
\label{LUBT}%
\end{equation}
that preserve the commutation relations of Eq.~(\ref{CCR}). It is easy to show
that such a preservation occurs when the matrix $\mathbf{S}$ is symplectic,
i.e., when it satisfies%
\begin{equation}
\mathbf{S\Omega S}^{T}=\mathbf{\Omega}~. \label{Sympl_cond}%
\end{equation}
By applying the map of Eq.~(\ref{LUBT}) to the Weyl operator of
Eq.~(\ref{Weyl operator}), we find the corresponding transformations for the
Wigner representations. In particular, the arbitrary vector $\mathbf{x}$ of
the phase space $\mathcal{K}=(\mathbb{R}^{2n},\mathbf{\Omega})$ undergoes
exactly the same affine map as above%
\begin{equation}
(\mathbf{S},\mathbf{d}):\mathbf{x}\rightarrow\mathbf{Sx+d}\boldsymbol{~}.
\label{Affine_map2}%
\end{equation}

In other words, an arbitrary Gaussian unitary $\hat{U}_{\mathbf{S},\mathbf{d}%
}$ acting on the Hilbert space $\mathcal{H}$ of the system is equivalent to a
symplectic affine map $(\mathbf{S},\mathbf{d})$ acting on the corresponding
phase space $\mathcal{K}$. Notice that such a map is composed by two different
elements, i.e., the phase-space displacement $\mathbf{d}\in\mathbb{R}^{2n}$
which corresponds to a displacement operator $\hat{D}(\mathbf{d})$, and the
symplectic transformation $\mathbf{S}$ which corresponds to a canonical
unitary $\hat{U}_{\mathbf{S}}$ in the Hilbert space. In particular, the
phase-space displacement does not affect the second moments of the quantum
state since the CM is transformed by the simple congruence%
\begin{equation}
\mathbf{V\rightarrow SVS}^{T}~. \label{CM_congruence}%
\end{equation}

Fundamental properties of the bosonic states can be easily expressed via the
symplectic manipulation of their CM. In fact, according to the Williamson's
theorem~\cite{Williamson,Arnold,Alex}, any CM $\mathbf{V}$ can be diagonalized
by a symplectic transformation. This means that there always exists a
symplectic matrix $\mathbf{S}$ such that
\begin{equation}
\mathbf{\mathbf{S}V\mathbf{S}}^{T}=\mathrm{diag}(\nu_{1},\nu_{1},\cdots
,\nu_{n},\nu_{n})~, \label{William_DEC}%
\end{equation}
where the set $\{\nu_{1},\cdots,\nu_{n}\}$ is called the \emph{symplectic
spectrum} and satisfies $\prod\nolimits_{k=1}^{n}\nu_{k}=\sqrt{\det\mathbf{V}%
}$ (since $\det\mathbf{\mathbf{S}}=1$ for symplectic $\mathbf{S}$). By
applying the symplectic diagonalization of Eq.~(\ref{William_DEC}) to
Eq.~(\ref{BONA_FIDE}), one can write the uncertainty principle in the simple
form of~\cite{RMP}
\begin{equation}
\nu_{k}\geq1~~\text{and~}~\mathbf{V}>0~. \label{Heis_spectrum}%
\end{equation}

\subsection{Gaussian channels and canonical forms\label{CFormsSec}}

A single-mode bosonic channel is a completely positive trace preserving (CPTP)
map $\mathcal{E}:\rho\rightarrow\mathcal{E}(\rho)$ acting on the density
matrix $\rho$\ of a single bosonic mode. In particular, it is Gaussian
($\mathcal{E}=\mathcal{G}$) if it transforms Gaussian states into Gaussian
states. The general form of a single-mode Gaussian channel can be expressed by
the following transformation of the characteristic
function~\cite{HolevoCanonical}
\begin{equation}
\mathcal{G}:\chi(\boldsymbol{\xi})\rightarrow\chi(\mathbf{T}\boldsymbol{\xi
})\exp\left(  -\tfrac{1}{2}\boldsymbol{\xi}^{T}\mathbf{N}\boldsymbol{\xi
}+i\mathbf{d}^{T}\boldsymbol{\xi}\right)  ~, \label{Gaussian_Map}%
\end{equation}
where $\mathbf{d}\in\mathbb{R}^{2}$ is a displacement, while $\mathbf{T}$ and
$\mathbf{N}$ are $2\times2$ real matrices, with $\mathbf{N}^{T}=\mathbf{N}%
\geq0$ and
\begin{equation}
\det\mathbf{N}\geq\left(  \det\mathbf{T}-1\right)  ^{2}. \label{bona_fide_N}%
\end{equation}
These are the transmission matrix $\mathbf{T}$ and the noise matrix
$\mathbf{N}$. At the level of the first two statistical moments, the
transformation of Eq.~(\ref{Gaussian_Map}) takes the simple form%
\begin{equation}
\mathbf{\bar{x}\rightarrow\mathbf{T}\bar{x}+d},~~\mathbf{V}\rightarrow
\mathbf{TVT}^{T}+\mathbf{N.} \label{mapGG}%
\end{equation}

\begin{figure}[ptbh]
\vspace{-0cm}
\par
\begin{center}
\includegraphics[width=0.48\textwidth] {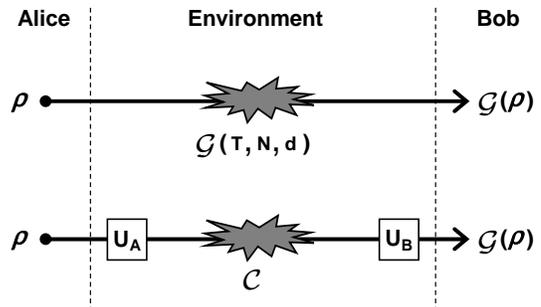}
\end{center}
\par
\vspace{-2.0cm}\caption{Reduction of a one-mode Gaussian channel $\mathcal{G}$
to its corresponding canonical form $\mathcal{C}$ by means of input-output
Gaussian unitaries $\hat{U}_{A}$ and $\hat{U}_{B}$.}%
\label{CanFormsPic}%
\end{figure}

Any single-mode Gaussian channel $\mathcal{G}=\mathcal{G}[\mathbf{T}%
,\mathbf{N},\mathbf{d}]$ can be transformed into a simpler \emph{canonical
form}~\cite{HolevoCanonical,Caruso,HolevoVittorio} via unitary transformations
at the input and the output (see Fig.~\ref{CanFormsPic}). In fact, for any
physical $\mathcal{G}$ there are (non-unique) finite-energy Gaussian unitaries
$\hat{U}_{A}$ and $\hat{U}_{B}$ such that%
\begin{equation}
\mathcal{G}(\rho)=\hat{U}_{B}\left[  \mathcal{C}(\hat{U}_{A}\rho\hat{U}%
_{A}^{\dagger})\right]  \hat{U}_{B}^{\dagger}~, \label{unitaryEQ}%
\end{equation}
where the canonical form $\mathcal{C}$ is the CPTP\ map%
\begin{equation}
\mathcal{C}:\chi(\boldsymbol{\xi})\rightarrow\chi(\mathbf{T}_{c}%
\boldsymbol{\xi})\exp\left(  -\tfrac{1}{2}\boldsymbol{\xi}^{T}\mathbf{N}%
_{c}\boldsymbol{\xi}\right)  ~, \label{Canonical_channel}%
\end{equation}
characterized by zero displacement ($\mathbf{d}=\mathbf{0}$) and diagonal
matrices $\mathbf{T}_{c}$ and $\mathbf{N}_{c}$. 

Depending on the values of the symplectic invariants $\det\mathbf{T}$,
rank($\mathbf{T}$) and rank($\mathbf{N}$), we have six different expressions
for the diagonal matrices $\mathbf{T}_{c},\mathbf{N}_{c}$ and, therefore, six
inequivalent classes of canonical forms $\mathcal{C}=\mathcal{C}%
[\mathbf{T}_{c},\mathbf{N}_{c}]$, which are denoted by $A_{1},A_{2}%
,B_{1},B_{2},C$ and $D$. From Ref.~\cite{HolevoVittorio} we report the
classification of these forms in Table \ref{TABLEcomp}, where $\mathbf{Z}%
:=\mathrm{diag}(1,-1)$, $\mathbf{I}$ the identity matrix, and $\mathbf{0}$ the
zero matrix. In this table $\tau:=\det\mathbf{T}$ is the (generalized)
transmissivity, while $\bar{n}\geq0$ is the thermal number of the environment
and $\xi\geq0$ is additive noise~\cite{notation}.\begin{table}[ptbh]
\centering
\[%
\begin{tabular}
[c]{c|c|c||c||c|c}%
$\tau:=\det\mathbf{T}$ & rk($\mathbf{T}$) & rk($\mathbf{N}$) & class &
$\mathbf{T}_{c}$ & $\mathbf{N}_{c}$\\\hline
$0$ & $0$ & $2$ & $A_{1}$ & $\mathbf{0}$ & $(2\bar{n}+1)\mathbf{I}$\\
$0$ & $1$ & $2$ & $A_{2}$ & $\frac{\mathbf{I}+\mathbf{Z}}{2}$ & $(2\bar
{n}+1)\mathbf{I}$\\
$1$ & $2$ & $1$ & $B_{1}$ & $\mathbf{I}$ & $\frac{\mathbf{I}-\mathbf{Z}}{2}$\\
$1$ & $2$ & $\neq1$ & $B_{2}$ & $\mathbf{I}$ & $\xi\mathbf{I}$\\
$0<\tau\neq1$ & $2$ & $2$ & $C$ & $\sqrt{\tau}\mathbf{I}$ & $\left\vert
1-\tau\right\vert (2\bar{n}+1)\mathbf{I}$\\
$\tau<0$ & $2$ & $2$ & $D$ & $\sqrt{-\tau}\mathbf{Z}$ & $(1-\tau)(2\bar
{n}+1)\mathbf{I}$%
\end{tabular}
\]
\caption{Classification of canonical
forms~\cite{HolevoCanonical,Caruso,HolevoVittorio}.}%
\label{TABLEcomp}%
\end{table}

Let us also introduce the symplectic invariant%
\begin{equation}
r:=\frac{\text{rank}(\mathbf{T})~\text{rank}(\mathbf{N})}{2}~,
\end{equation}
that we call the \emph{rank} of the Gaussian channel~\cite{formsREF,RMP}.
Then, every class is simply determined by the pair $\{\tau,r\}$ according to
the refined Table~\ref{TABLEcomp2}. Note that classes $B_{2}$ and $C$ have
been divided into subclasses. In fact, class $B_{2}$\ includes the identity
channel (for $r=0$), while class $C$ describes an attenuator (amplifier)
channel for $0<\tau<1$ ($\tau>1$). In common terminology the forms $A_{1}$,
$B_{2}$ and $C$ are known as phase-insensitive, because they act symmetrically
on the two input quadratures. By contrast, the forms $A_{2}$, $B_{1}$ and $D$
(conjugate of the amplifier) are all phase-sensitive. The form $B_{2}$ is an
additive form. In fact it is also known as the additive-noise Gaussian
channel, which is a direct generalization of the classical Gaussian channel in
the quantum setting.\begin{table}[ptbh]
\centering%
\[%
\begin{tabular}
[c]{c|c||c|c||c|c||c}%
$\tau$ & $~r~$ & class & sub & $\mathbf{T}_{c}$ & $\mathbf{N}_{c}$ &
$\mathcal{C}[\tau,r,\bar{n}]$\\\hline
$0$ & $0$ & $A_{1}$ &  & $\mathbf{0}$ & $(2\bar{n}+1)\mathbf{I}$ &
$\mathcal{C}[0,0,\bar{n}]$\\
$0$ & $1$ & $A_{2}$ &  & $\frac{\mathbf{I}+\mathbf{Z}}{2}$ & $(2\bar
{n}+1)\mathbf{I}$ & $\mathcal{C}[0,1,\bar{n}]$\\
$1$ & $1$ & $B_{1}$ &  & $\mathbf{I}$ & $\frac{\mathbf{I}-\mathbf{Z}}{2}$ &
$\mathcal{C}[1,1,0]$\\
$1$ & $2$ & $B_{2}$ & $\neq$Id & $\mathbf{I}$ & $\xi\mathbf{I}$ &
$\mathcal{C}[1,2,\xi]$\\
$1$ & $0$ & $B_{2}$ & Id & $\mathbf{I}$ & $\mathbf{0}$ & $\mathcal{C}%
[1,0,0]$\\
$]0,1[$ & $2$ & $C$ & Att & $\sqrt{\tau}\mathbf{I}$ & $(1-\tau)(2\bar
{n}+1)\mathbf{I}$ & $\mathcal{C}[\tau,2,\bar{n}]$\\
$>1$ & $2$ & $C$ & Amp & $\sqrt{\tau}\mathbf{I}$ & $(\tau-1)(2\bar
{n}+1)\mathbf{I}$ & $\mathcal{C}[\tau,2,\bar{n}]$\\
$<0$ & $2$ & $D$ &  & $\sqrt{-\tau}\mathbf{Z}$ & $(1-\tau)(2\bar
{n}+1)\mathbf{I}$ & $\mathcal{C}[\tau,2,\bar{n}]$%
\end{tabular}
\ \ \
\]
\caption{Refined classification of canonical forms.}%
\label{TABLEcomp2}%
\end{table}

\subsection{Single-mode dilation of a canonical form\label{PRSection}}

\begin{figure}[ptbh]
\vspace{-0cm}
\par
\begin{center}
\includegraphics[width=0.48\textwidth] {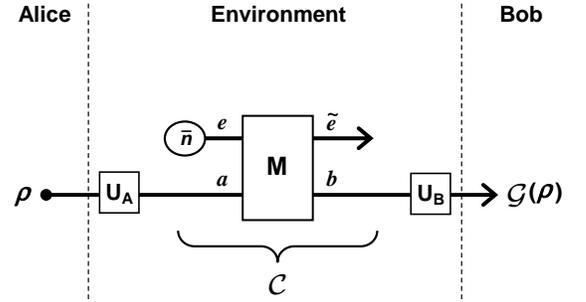}
\end{center}
\par
\vspace{-2.0cm}\caption{Single-mode physical representation of a non-additive
canonical form $\mathcal{C}$ (all forms but $B_{2}$). This is also the
physical representation of a non-additive Gaussian channel up to the
input-output unitaries $\hat{U}_{A}$ and $\hat{U}_{B}$.}%
\label{PRnonadd}%
\end{figure}

All the non-additive forms $\mathcal{C}[\tau,r,\bar{n}]$ admit a simple
single-mode physical representation where the degrees of freedom $\hat{x}%
_{a}^{T}:=(\hat{q}_{a},\hat{p}_{a})$ of the input bosonic mode
\textquotedblleft$a$\textquotedblright\ unitarily interacts with the degrees
of freedom $\hat{x}_{e}^{T}:=(\hat{q}_{e},\hat{p}_{e})$ of a single
environmental bosonic mode \textquotedblleft$e$\textquotedblright\ described
by a mixed state $\rho_{e}$ \cite{Caruso,HolevoVittorio}\ (see
Fig.~\ref{PRnonadd}). In particular, such a physical representation can always
be chosen to be Gaussian. This means that $\mathcal{C}[\tau,r,\bar{n}]$ can be
represented by a canonical unitary $\hat{U}_{ae}$ mixing the input state
$\rho_{a}$ with a thermal state $\rho_{e}(\bar{n})$, i.e.,%
\begin{equation}
\mathcal{C}:\rho_{a}\rightarrow\mathcal{C}(\rho_{a})=\text{Tr}_{e}\left\{
\hat{U}_{ae}\left[  \rho_{a}\otimes\rho_{e}(\bar{n})\right]  \hat{U}%
_{ae}^{\dagger}\right\}  , \label{PR}%
\end{equation}
where
\begin{equation}
\hat{U}_{ae}\left(
\begin{array}
[c]{c}%
\hat{x}_{a}\\
\hat{x}_{e}%
\end{array}
\right)  \hat{U}_{ae}^{\dagger}=\mathbf{M}\left(
\begin{array}
[c]{c}%
\hat{x}_{a}\\
\hat{x}_{e}%
\end{array}
\right)  ,
\end{equation}
with $\mathbf{M}$ symplectic and $\rho_{e}(\bar{n})$ is a thermal state with
CM $\mathbf{V}_{e}(\bar{n})=(2\bar{n}+1)\mathbf{I}$ (see Fig.~\ref{PRnonadd}).

In fact, by writing $\mathbf{M}$ in the blockform%
\begin{equation}
\mathbf{M}=\left(
\begin{array}
[c]{cc}%
\mathbf{m}_{1} & \mathbf{m}_{2}\\
\mathbf{m}_{3} & \mathbf{m}_{4}%
\end{array}
\right)  , \label{Blockform}%
\end{equation}
so that
\begin{align}
\hat{x}_{a}  &  \rightarrow\hat{x}_{b}:=\mathbf{m}_{1}\hat{x}_{a}%
+\mathbf{m}_{2}\hat{x}_{e}~,\\
\hat{x}_{e}  &  \rightarrow\hat{x}_{\tilde{e}}:=\mathbf{m}_{3}\hat{x}%
_{a}+\mathbf{m}_{4}\hat{x}_{e}~, \label{LUBT_e_eprime}%
\end{align}
one finds that Eq.~(\ref{PR}) corresponds to the following input-output
transformation for the characteristic function%
\begin{equation}
\chi_{a}(\boldsymbol{\xi})\rightarrow\chi_{a}(\mathbf{m}_{1}^{T}%
\boldsymbol{\xi})\exp\left[  -\tfrac{1}{2}(2\bar{n}+1)\left\vert
\mathbf{m}_{2}^{T}\boldsymbol{\xi}\right\vert ^{2}\right]  .
\label{Characteristic_PR}%
\end{equation}

Then, by setting $\mathbf{m}_{1}^{T}=\mathbf{T}_{c}$ and%
\begin{equation}
\mathbf{m}_{2}^{T}=\sqrt{\frac{\mathbf{N}_{c}}{2\bar{n}+1}}\mathbf{O}%
,~~\mathbf{O}^{T}=\mathbf{O}^{-1}, \label{A_Atilde}%
\end{equation}
one easily verifies that Eq.~(\ref{Characteristic_PR}) has the form of
Eq.~(\ref{Canonical_channel}), where the bona fide condition of
Eq.~(\ref{bona_fide_N}) is assured by the symplectic nature of $\mathbf{M}%
$~\cite{NoteSympl}. In Eq.~(\ref{A_Atilde}) the orthogonal transformation
$\mathbf{O}$ is chosen in a way to preserve the symplectic condition for
$\mathbf{M}$. Such a condition also restricts the possible forms of the
remaining blocks $\mathbf{m}_{3}$ and $\mathbf{m}_{4}$, which can be fixed up
to a canonical local unitary. 

Altogether, any non-additive canonical form $\mathcal{C}[\tau,r,\bar{n}]$ can
be described by a single-mode physical representation $\{\mathbf{M}%
(\tau,r),\rho_{e}(\bar{n})\}$ where the type of symplectic transformation
$\mathbf{M}(\tau,r)$ is determined by its class $\{\tau,r\}$\ while the
thermal noise $\bar{n}$ only characterizes the environmental state. From the
point of view of the second order statistical moments, the CM $\mathbf{V}_{a}$
of an input state $\rho_{a}$\ undergoes the transformation
\begin{equation}
\mathbf{V}_{a}\rightarrow\mathrm{Tr}_{e}\left\{  \mathbf{M}_{ae}%
(\tau,r)\left[  \mathbf{V}_{a}\mathbf{\oplus}(2\bar{n}+1)\mathbf{I}%
_{e}\right]  \mathbf{M}_{ae}(\tau,r)^{T}\right\}  ,
\end{equation}
where the partial trace $\mathrm{Tr}_{e}$ must be interpreted as deletion of
rows and columns associated with mode $e$.

In particular, one has the following symplectic matrices for the various
forms~\cite{HolevoVittorio}%
\begin{equation}
\mathbf{M}(0<\tau<1,2)=\mathbf{M}(C)=\left(
\begin{array}
[c]{cc}%
\sqrt{\tau}\mathbf{I} & \sqrt{1-\tau}\mathbf{I}\\
-\sqrt{1-\tau}\mathbf{I} & \sqrt{\tau}\mathbf{I}%
\end{array}
\right)  ,
\end{equation}
describing a beam-splitter,%
\begin{equation}
\mathbf{M}(\tau>1,2)=\mathbf{M}(C)=\left(
\begin{array}
[c]{cc}%
\sqrt{\tau}\mathbf{I} & \sqrt{\tau-1}\mathbf{Z}\\
\sqrt{\tau-1}\mathbf{Z} & \sqrt{\tau}\mathbf{I}%
\end{array}
\right)  ,
\end{equation}
describing an amplifier,%
\begin{equation}
\mathbf{M}(\tau<0,2)=\mathbf{M}(D)=\left(
\begin{array}
[c]{cc}%
\sqrt{-\tau}\mathbf{Z} & \sqrt{1-\tau}\mathbf{I}\\
-\sqrt{1-\tau}\mathbf{I} & -\sqrt{-\tau}\mathbf{Z}%
\end{array}
\right)  ,
\end{equation}
describing the complementary of an amplifier. Finally~\cite{HolevoVittorio}%
\begin{align}
\mathbf{M}(0,0)  &  =\mathbf{M}(A_{1})=\left(
\begin{array}
[c]{cc}%
\mathbf{0} & \mathbf{I}\\
\mathbf{I} & \mathbf{0}%
\end{array}
\right)  ,\\
\mathbf{M}(0,1)  &  =\mathbf{M}(A_{2})=\left(
\begin{array}
[c]{cc}%
\frac{\mathbf{I+Z}}{2} & \mathbf{I}\\
\mathbf{I} & \frac{\mathbf{Z-I}}{2}%
\end{array}
\right)  ,\\
\mathbf{M}(1,1)  &  =\mathbf{M}(B_{1})=\left(
\begin{array}
[c]{cc}%
\mathbf{I} & \frac{\mathbf{I+Z}}{2}\\
\frac{\mathbf{I-Z}}{2} & \mathbf{-I}%
\end{array}
\right)  .
\end{align}

\subsection{Asymptotic dilation of the additive $B_{2}$ form\label{B2section}}

The additive-noise Gaussian channel or $B_{2}$ canonical form $\mathcal{C}%
[1,2,\xi]$ can be dilated into a two-mode environment~\cite{HolevoVittorio}.
Another possibility is to describe this form by means of an asymptotic
single-mode dilation. In fact, consider the dilation of the attenuator
channel, which is a beam-splitter $\hat{U}_{ae}^{\text{BS}}(\tau)$ with
transmissivity $\tau$ coupling the input mode $a$ with an environmental mode
$e$ prepared in a thermal state $\rho_{e}(\bar{n})$ with $\bar{n}$ mean
photons. In this dilation, let us consider a thermal state with $\bar{n}%
_{\xi,\tau}:=[\xi(1-\tau)^{-1}-1]/2$ so that we realize $(1-\tau)(2\bar
{n}+1)=\xi$. Then, taking the limit for $\tau\rightarrow1$ (so that $\bar
{n}\rightarrow+\infty$), we represent the $B_{2}$ canonical form as%
\begin{equation}
\mathcal{C}[1,2,\xi](\rho_{a})=\lim_{\tau\rightarrow1}\text{Tr}_{e}\left\{
\hat{U}_{ae}^{\text{BS}}(\tau)\left[  \rho_{a}\otimes\rho_{e}(\bar{n}%
_{\xi,\tau})\right]  \hat{U}_{ae}^{\text{BS}}(\tau)^{\dagger}\right\}  .
\end{equation}
In fact it is clear that, in this way, we may realize the asymptotic
transformations $\mathbf{x}\rightarrow\mathbf{x}$ and $\mathbf{V}%
\longrightarrow\mathbf{V+}\xi\mathbf{I}$.

\section{Convergence of CV teleportation}

\subsection{Braunstein-Kimble teleportation protocol}

Let us review the BK protocol for CV quantum
teleportation~\cite{Tele2,telereview}. Alice and Bob share a resource state
which is a two-mode squeezed vacuum (TMSV) state $\Phi_{AB}^{\mu}$. Recall
that this is a zero-mean Gaussian state with CM~\cite{RMP}%
\begin{equation}
\mathbf{V}^{\mu}=\left(
\begin{array}
[c]{cc}%
\mu\mathbf{I} & \sqrt{\mu^{2}-1}\mathbf{Z}\\
\sqrt{\mu^{2}-1}\mathbf{Z} & \mu\mathbf{I}%
\end{array}
\right)  . \label{TMSVstate}%
\end{equation}
Here the variance parameter $\mu$ determines both the squeezing (or
entanglement) and the energy associated with the state. In particular, we may
write $\mu=2\bar{n}+1$, where $\bar{n}$ is the mean number of photons in each
mode, $A$ (for Alice)\ and $B$ (for Bob).

Then, Alice has an input bipartite state $\rho_{Ra}$, where $R$ is an
arbitrary multimode system while $a$ is a single mode that she wants to
teleport to Bob. To teleport, she combines modes $a$ and $A$ in a joint CV
Bell detection, whose complex outcome $\alpha$ is classically communicated to
Bob (this can be realized by a balanced beam splitter followed by two
conjugate homodyne detectors~\cite{telereview}). Finally, Bob applies a
displacement $D(-\alpha)$\ on his mode $B$, so that the output state
$\rho_{RB}$ is the teleported version $\rho_{Ra}^{\mu}$ of the input
$\rho_{Ra}$.

One has perfect teleportation in the limit of infinite squeezing $\mu$. In
other words, for any input state $\rho_{Ra}$ (with finite energy) we may write
the trace norm limit
\begin{equation}
\lim_{\mu\rightarrow\infty}\left\Vert \rho_{Ra}^{\mu}-\rho_{Ra}\right\Vert =0,
\end{equation}
or equivalently, we may write%
\begin{equation}
\lim_{\mu\rightarrow\infty}F(\rho_{Ra}^{\mu},\rho_{Ra})=1,
\end{equation}
where $F(\rho,\sigma):=\mathrm{Tr}\sqrt{\sqrt{\sigma}\rho\sqrt{\sigma}}$ is
the Bures fidelity. This is a well known result which has been proven in
Ref.~\cite{Tele2}.

\subsection{Strong convergence of CV\ teleportation}

\begin{figure*}[ptbh]
\vspace{-3cm}
\par
\begin{center}
\includegraphics[width=0.95\textwidth] {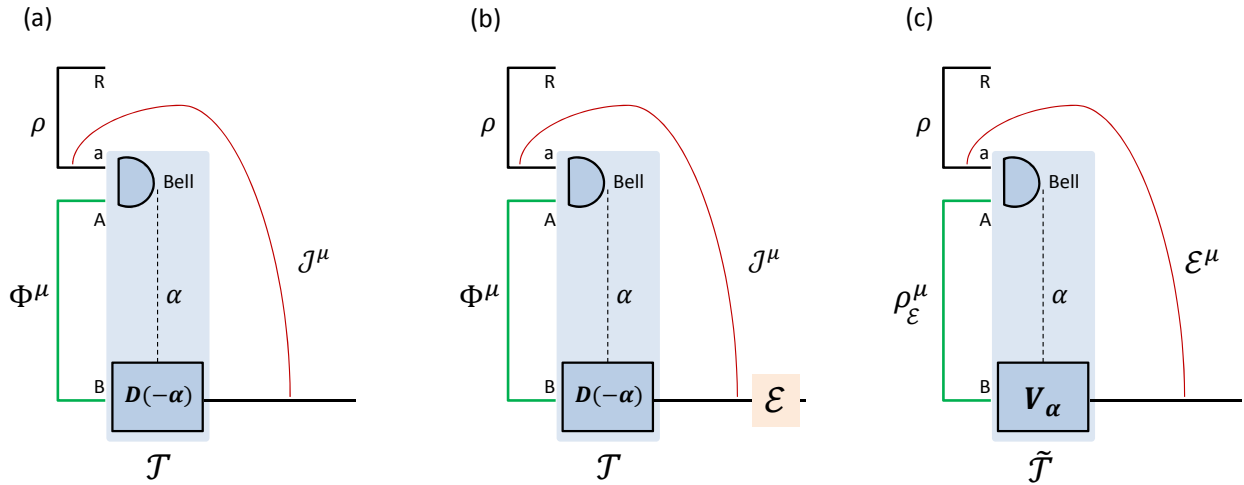}
\end{center}
\par
\vspace{-3cm}\caption{BK protocol and teleportation simulation of bosonic
channels. (a)~We depict the BK protocol, where Alice's mode $a$ of a bipartite
input state $\rho_{Ra}$ is teleported to Bob's mode $B$. This is performed by
detecting mode $a$ along with mode $A$ of a TMSV\ state $\Phi^{\mu}$ via a
Bell measurement. The complex output $\alpha$ is then used to apply the
conditional displacement $D(-\alpha)$ on mode $B$. The output state $\rho
_{RB}$ is an approximate version $\rho_{Ra}^{\mu}$ of the input state
$\rho_{Ra}$. This output can be written as in Eq.~(\ref{eqBKfig}) where
$\mathcal{I}^{\mu}$ is the BK channel and $\mathcal{T}_{aAB}$ is the overall
teleportation LOCC (Bell detection plus conditional displacements).
(b)~Consider a bosonic channel $\mathcal{E}$ applied to the output mode $B$,
so that we may define the composite channel $\mathcal{E}^{\mu}=\mathcal{E}%
\circ\mathcal{I}^{\mu}$ as in Eq.~(\ref{monokk}). (c)~If $\mathcal{E}$ is
teleportation-covariant, we may write its teleportation simulation
$\mathcal{E}^{\mu}$ as in Eq.~(\ref{teleBOSONIC}) where the
modified\ teleportation LOCC $\mathcal{\tilde{T}}$ (Bell detection and unitary
corrections $V_{\alpha}$) is applied to the input and the quasi-Choi state
$\rho_{\mathcal{E}}^{\mu}:=\mathcal{I}\otimes\mathcal{E}(\Phi^{\mu})$.}%
\label{BKprotocol}%
\end{figure*}

Let us denote by $\mathcal{T}$ the overall LOCC associated with the
BK\ protocol, as in \ Fig.~\ref{BKprotocol}(a). The application of this LOCC
onto a finite-energy TMSV state $\Phi^{\mu}$ generates a teleportation channel
$\mathcal{I}^{\mu}$ which is not the bosonic identity channel $\mathcal{I}$
but a point-wise (local) approximation of $\mathcal{I}$. In other words, for
any (energy-bounded) input state $\rho_{Ra}$, we may consider the output
\begin{equation}
\rho_{Ra}^{\mu}:=\mathcal{I}_{R}\otimes\mathcal{I}_{a}^{\mu}(\rho
_{Ra})=\mathcal{I}_{R}\otimes\mathcal{T}_{aAB}(\rho_{Ra}\otimes\Phi_{AB}^{\mu
}), \label{eqBKfig}%
\end{equation}
and write the trace-norm limit
\begin{equation}
\lim_{\mu\rightarrow\infty}\left\Vert \mathcal{I}_{R}\otimes\mathcal{I}%
_{a}^{\mu}(\rho_{Ra})-\rho_{Ra}\right\Vert =0. \label{pointwise}%
\end{equation}
It is clear that this point-wise limit immediately implies the convergence in
the strong topology~\cite{Tele2}%
\begin{equation}
\sup_{\rho_{Ra}}\lim_{\mu\rightarrow\infty}\left\Vert \mathcal{I}_{R}%
\otimes\mathcal{I}_{a}^{\mu}(\rho_{Ra})-\rho_{Ra}\right\Vert =0.
\label{strongBK}%
\end{equation}
Similarly, we may introduce the Bures distance~\cite{qfi1}
\begin{equation}
d_{\text{B}}(\rho,\sigma):=\sqrt{2[1-F(\rho,\sigma)]},
\end{equation}
and write the previous limit as%
\begin{equation}
\sup_{\rho_{Ra}}\lim_{\mu\rightarrow\infty}d_{\text{B}}[\mathcal{I}_{R}%
\otimes\mathcal{I}_{a}^{\mu}(\rho_{Ra}),\rho_{Ra}]=0.
\end{equation}

\begin{remark}
Let us stress that the strong convergence of the BK protocol is
known since 1998. It is well-known that, for any given
energy-constrained input state, if we send the squeezing of the
resource state (TMSV state) to infinite, then we can perfectly
teleport the input state. In Eqs.~(4) and~(8) of
Ref.~\cite{Tele2}, there is a convolution between the Wigner
function $W_{\text{in}}$ of an arbitrary normalized input state
and the Gaussian kernel $G_{\sigma}$, where $\sigma$ goes to zero
for increasing squeezing $r$ (and ideal homodyne detectors).
Taking the limit for large $r$, the teleportation fidelity goes to
$1$ as we can also see from Eq. (11)\ of Ref.~\cite{Tele2}. This
is just a standard delta-like limit that does not really need
explicit steps to be shown and fully provides the (strong)
convergence of the BK protocol.
\end{remark}

\subsection{Bounded-uniform convergence of CV teleportation}

Consider an energy-constrained alphabet of states
\begin{equation}
\mathcal{D}_{N}:=\{\rho_{Ra}~|~\mathrm{Tr}(\hat{N}\rho_{Ra})\leq N\},
\label{alphabetBOUNDED}%
\end{equation}
where $\hat{N}$ is the total number operator associated with the input mode
$a$ and the reference modes $R$. Then, we define an energy-constrained diamond
distance~\cite{PLOB,TQC} between two arbitrary bosonic channels $\mathcal{E}$
and $\mathcal{E}^{\prime}$, as%
\begin{equation}
\left\Vert \mathcal{E}-\mathcal{E}^{\prime}\right\Vert _{\diamond N}%
:=\sup_{\rho_{Ra}\in\mathcal{D}_{N}}\left\Vert \mathcal{I}_{R}\otimes
\mathcal{E}_{a}(\rho_{Ra})-\mathcal{I}_{R}\otimes\mathcal{E}_{a}^{\prime}%
(\rho_{Ra})\right\Vert ~. \label{boundedDIAMOND}%
\end{equation}
See also Ref.~\cite{MaximNORM,WinterNORM}\ for an alternate definition of
energy-constrained diamond norm. It is easy to show that, for any finite
energy $N$, one may write~\cite{PLOB}
\begin{equation}
\lim_{\mu\rightarrow\infty}\left\Vert \mathcal{I}^{\mu}-\mathcal{I}\right\Vert
_{\diamond N}=0, \label{BKuniform}%
\end{equation}
so that the BK\ channel $\mathcal{I}^{\mu}$ converges to the identity channel
in the bounded-uniform topology. In fact, this comes from the point-wise limit
in Eq.~(\ref{pointwise}) combined with the fact that $\mathcal{D}_{N}$ is a
compact set~\cite{HolevoCOMPACT,HolBook,Werner}.

\subsection{Non-uniform convergence of CV\ teleportation\label{nonUNIsec}}

Can we relax the energy constraint $N$ in Eq.~(\ref{BKuniform})? The answer is
no. As already discussed in Ref.~\cite{TQC}, we have
\begin{equation}
\lim_{\mu\rightarrow\infty}\left\Vert \mathcal{I}^{\mu}-\mathcal{I}\right\Vert
_{\diamond}=2, \label{nonUNI}%
\end{equation}
where
\begin{align}
\left\Vert \mathcal{E}-\mathcal{E}^{\prime}\right\Vert _{\diamond}  &
=\lim_{N\rightarrow\infty}\left\Vert \mathcal{E}-\mathcal{E}^{\prime
}\right\Vert _{\diamond N}\\
&  =\sup_{\rho_{Ra}}\left\Vert \mathcal{I}_{R}\otimes\mathcal{E}_{a}(\rho
_{Ra})-\mathcal{I}_{R}\otimes\mathcal{E}_{a}^{\prime}(\rho_{Ra})\right\Vert
\end{align}
is the standard diamond distance. In fact, Ref.~\cite{TQC} provided a simple
proof that the BK\ protocol does not uniformly converge to the identity
channel. For this proof, it is sufficient to take the input state to be a TMSV
state $\Phi^{\tilde{\mu}}$ with diverging energy $\tilde{\mu}$. Then,
Eq.~(\ref{nonUNI}) is implied by the fact that, for any $\mu$-energy
BK\ protocol, we have
\begin{equation}
\lim_{\tilde{\mu}\rightarrow\infty}\left\Vert \mathcal{I}_{R}\otimes
\mathcal{I}_{a}^{\mu}(\Phi_{Ra}^{\tilde{\mu}})-\Phi_{Ra}^{\tilde{\mu}%
}\right\Vert =2, \label{Toshow}%
\end{equation}
which is equivalent to $\left\Vert \mathcal{I}^{\mu}-\mathcal{I}\right\Vert
_{\diamond}=2$ for any $\mu$.

In order to show Eq.~(\ref{Toshow}) we directly report the steps given in
Ref.~\cite{TQC} but adapted to our different notation. The first observation
is that, when applied to an energy-constrained quantum state (i.e., a
\textquotedblleft point\textquotedblright), the $\mu$-energy BK channel
$\mathcal{I}^{\mu}$ is locally equivalent to an additive-noise Gaussian
channel (form $B_{2}$) with added noise
\begin{equation}
\xi=2[\mu-\sqrt{\mu^{2}-1}]~. \label{addedNOISE}%
\end{equation}
For instance, see Refs.~\cite{GerLimited,next3}. Then, from the
CM\ $\mathbf{V}^{\tilde{\mu}}$ of $\Phi_{Ra}^{\tilde{\mu}}$, it is easy to
compute the CM\ of the output state $\rho_{Ra}^{\mu,\tilde{\mu}}%
:=\mathcal{I}_{R}\otimes\mathcal{I}_{a}^{\mu}(\Phi_{Ra}^{\tilde{\mu}})$
yielding%
\begin{equation}
\mathbf{V}^{\mu,\tilde{\mu}}=\left(
\begin{array}
[c]{cc}%
\tilde{\mu}\mathbf{I} & \sqrt{\tilde{\mu}^{2}-1}\mathbf{Z}\\
\sqrt{\tilde{\mu}^{2}-1}\mathbf{Z} & (\tilde{\mu}+\xi)\mathbf{I}%
\end{array}
\right)  .
\end{equation}
Using the formula for the quantum fidelity between arbitrary Gaussian
states~\cite{banchiPRL2015}, we compute%
\begin{gather}
F(\tilde{\mu},\mu):=F(\rho_{Ra}^{\mu,\tilde{\mu}},\Phi_{Ra}^{\tilde{\mu}%
})\label{fidLIMIT}\\
=\frac{1}{\sqrt[4]{1-4\tilde{\mu}\left[  \sqrt{4\mu^{2}-1}+\tilde{\mu}%
-2\mu(1+4\mu\tilde{\mu}-2\tilde{\mu}\sqrt{4\mu^{2}-1})\right]  }}.\nonumber
\end{gather}
Here we notice the expansion $F(\tilde{\mu},\mu)\simeq O(\tilde{\mu}^{-1/2})$
at any fixed $\mu$. Now using the Fuchs-van de Graaf relations~\cite{Fuchs}
\begin{equation}
2[1-F(\rho,\sigma)]\leq\left\Vert \rho-\sigma\right\Vert \leq2\sqrt
{1-F(\rho,\sigma)^{2}}, \label{FuchsGraaf}%
\end{equation}
we get, for any finite $\mu$, the following expansion
\begin{equation}
\left\Vert \rho_{Ra}^{\mu,\tilde{\mu}}-\Phi_{Ra}^{\tilde{\mu}}\right\Vert
\geq2-O(\tilde{\mu}^{-1/2}),
\end{equation}
which implies Eq.~(\ref{Toshow}).

Here it is important to observe the radically different behavior of the
teleportation protocol with respect to exchanging the limits in the energy
$\mu$ of the resource state $\Phi_{AB}^{\mu}$ and in the energy $\tilde{\mu}$
of the input state $\Phi_{Ra}^{\tilde{\mu}}$. In fact, by taking the limit in
$\mu$ \emph{before} the one in $\tilde{\mu}$ in Eq.~(\ref{fidLIMIT}), we get
\begin{equation}
F(\tilde{\mu},\mu)\simeq1-O(\mu^{-1}).
\end{equation}
Because of the non-commutation between these two limits%
\begin{equation}
\lim_{\mu}\left[  \lim_{\tilde{\mu}}F(\tilde{\mu},\mu)\right]  \neq
\lim_{\tilde{\mu}}\left[  \lim_{\mu}F(\tilde{\mu},\mu)\right]  ,
\end{equation}
we have a difference between the strong convergence in Eq.~(\ref{strongBK})
and the uniform non-convergence in Eq.~(\ref{nonUNI}). This also means that
joint limits such as
\begin{equation}
\lim_{\mu,\tilde{\mu}}F(\tilde{\mu},\mu),~~~\underset{\mu,\tilde{\mu}}%
{\lim\sup}~F(\tilde{\mu},\mu)
\end{equation}
are not defined. While this problem has been known since the early days of CV
teleportation, technical errors related to this issue can still be found in
recent literature (see the \textquotedblleft case study\textquotedblright%
\ discussed in Sec.~VII D).

\section{Teleportation simulation of bosonic channels}

The BK\ teleportation protocol is a fundamental tool for the simulation of
bosonic channels (not necessarily Gaussian). Consider a
teleportation-covariant bosonic channel $\mathcal{E}$~\cite{PLOB}. This means
that, for any random displacement $D(-\alpha)$, we may write
\begin{equation}
\mathcal{E}[D(-\alpha)\rho D(\alpha)]=V_{\alpha}\mathcal{E}(\rho)V_{\alpha
}^{\dagger}, \label{teleCOV}%
\end{equation}
where $V_{\alpha}$ is an output unitary. If this is the case, then the bosonic
channel $\mathcal{E}$ can be simulated by teleporting the input state with a
modified teleportation LOCC $\mathcal{\tilde{T}}$ over the (asymptotic) Choi
matrix $\rho_{\mathcal{E}}$ of the channel $\mathcal{E}$. In particular,
Eq.~(\ref{teleCOV}) is true for Gaussian channels, for which $V_{\alpha}$ is
just another displacement.

In order to correctly formulate this type of simulation, we need to start from
an imperfect finite-energy simulation and then take the asymptotic limit for
large energy. Therefore, let us consider a $\mu$-energy BK protocol
$(\mathcal{T},\Phi^{\mu})$ generating a BK channel $\mathcal{I}^{\mu}$\ at the
input of a bosonic channel $\mathcal{E}$. Let us consider the composite
channel
\begin{equation}
\mathcal{E}^{\mu}=\mathcal{E}\circ\mathcal{I}^{\mu}. \label{monokk}%
\end{equation}
As shown in Fig.~\ref{BKprotocol}(b), for any input state $\rho_{Ra}$, we may
write the output state as%
\begin{equation}
\mathcal{I}_{R}\otimes\mathcal{E}_{a}^{\mu}(\rho_{Ra})=\mathcal{I}_{R}%
\otimes\mathcal{E}_{B}\circ\mathcal{T}_{aAB}(\rho_{Ra}\otimes\Phi_{AB}^{\mu}).
\label{BKfirst}%
\end{equation}

If the bosonic channel $\mathcal{E}$ is teleportation covariant, then we can
swap it with the displacements $D(-\alpha)$, up to re-defining the
teleportation corrections as $V_{\alpha}$. On the one hand this changes the
teleportation LOCC $\mathcal{\tilde{T}}$, on the other hand the resource state
becomes a quasi-Choi state%
\begin{equation}
\rho_{\mathcal{E}}^{\mu}:=\mathcal{I}_{A}\circ\mathcal{E}_{B}(\Phi_{AB}^{\mu
}).
\end{equation}
Therefore, as depicted in Fig.~\ref{BKprotocol}(c), we may re-write the
teleportation simulation of the output as%
\begin{equation}
\mathcal{I}_{R}\otimes\mathcal{E}_{a}^{\mu}(\rho_{Ra})=\mathcal{I}_{R}%
\otimes\mathcal{\tilde{T}}_{aAB}\left[  \rho_{Ra}\otimes(\rho_{\mathcal{E}%
}^{\mu})_{AB}\right]  . \label{teleBOSONIC}%
\end{equation}

Now, using Eq.~(\ref{monokk}) and the monotonicity of the trace distance under
CPTP maps, we may write%
\begin{align}
&  \left\Vert \mathcal{I}_{R}\otimes\mathcal{E}_{a}^{\mu}(\rho_{Ra}%
)-\mathcal{I}_{R}\otimes\mathcal{E}_{a}(\rho_{Ra})\right\Vert \nonumber\\
&  =\left\Vert \mathcal{I}_{R}\otimes\mathcal{E}_{a}\circ\mathcal{I}_{a}^{\mu
}(\rho_{Ra})-\mathcal{I}_{R}\otimes\mathcal{E}_{a}\circ\mathcal{I}_{a}%
(\rho_{Ra})\right\Vert \nonumber\\
&  \leq\left\Vert \mathcal{I}_{R}\otimes\mathcal{I}_{a}^{\mu}(\rho_{Ra}%
)-\rho_{Ra}\right\Vert \overset{\mu\rightarrow\infty}{\rightarrow}0,
\end{align}
where we exploit Eq.~(\ref{pointwise}) in the last step. Therefore, for any
bipartite (energy-constrained) input state $\rho_{Ra}$, we may write the
point-wise limit%
\begin{equation}
\lim_{\mu\rightarrow\infty}\left\Vert \mathcal{I}_{R}\otimes\mathcal{E}%
_{a}^{\mu}(\rho_{Ra})-\mathcal{I}_{R}\otimes\mathcal{E}_{a}(\rho
_{Ra})\right\Vert =0. \label{limSTRONG}%
\end{equation}

\subsection{Strong convergence in the teleportation simulation of bosonic
channels}

The strong convergence in the simulation of (teleportation-covariant) bosonic
channels (not necessarily Gaussian) is an immediate consequence of the
point-wise limit in Eq.~(\ref{limSTRONG}). In fact, because
Eq.~(\ref{limSTRONG}) holds for any bipartite (energy-constrained) input state
$\rho_{Ra}$, we may write%
\begin{equation}
\sup_{\rho_{Ra}}\lim_{\mu\rightarrow\infty}\left\Vert \mathcal{I}_{R}%
\otimes\mathcal{E}_{a}^{\mu}(\rho_{Ra})-\mathcal{I}_{R}\otimes\mathcal{E}%
_{a}(\rho_{Ra})\right\Vert =0,
\end{equation}
or similarly in terms of the Bures distance
\begin{equation}
\sup_{\rho_{Ra}}\lim_{\mu\rightarrow\infty}d_{\text{B}}[\mathcal{I}_{R}%
\otimes\mathcal{E}_{a}^{\mu}(\rho_{Ra}),\mathcal{I}_{R}\otimes\mathcal{E}%
_{a}(\rho_{Ra})]=0.
\end{equation}
In\ other words, the teleportation simulation $\mathcal{E}^{\mu}$ of a bosonic
channel $\mathcal{E}$, strongly converges to it in the limit of large $\mu$.

\subsection{Bounded-uniform convergence in the teleportation simulation of
bosonic channels}

Consider now an energy constrained input alphabet $\mathcal{D}_{N}$ as in
Eq.~(\ref{alphabetBOUNDED}) and the energy-constrained diamond distance
defined in Eq.~(\ref{boundedDIAMOND}). Given an arbitrary
(teleportation-covariant) bosonic channel $\mathcal{E}$ and its teleportation
simulation $\mathcal{E}^{\mu}$ as in Eq.~(\ref{teleBOSONIC}), we define the
simulation error as~\cite{PLOB,TQC}%
\begin{equation}
\delta(\mu,N):=\left\Vert \mathcal{E}^{\mu}-\mathcal{E}\right\Vert _{\diamond
N}~. \label{already}%
\end{equation}
Because of the monotonicity of the trace-distance under CPTP maps, we may
certainly write%
\begin{align}
\delta(\mu,N)  &  =\sup_{\rho_{Ra}\in\mathcal{D}_{N}}\left\Vert \mathcal{I}%
_{R}\otimes\mathcal{E}_{a}^{\mu}(\rho_{Ra})-\mathcal{I}_{R}\otimes
\mathcal{E}_{a}(\rho_{Ra})\right\Vert \\
&  \leq\sup_{\rho_{Ra}\in\mathcal{D}_{N}}\left\Vert \mathcal{I}_{R}%
\otimes\mathcal{I}_{a}^{\mu}(\rho_{Ra})-\rho_{Ra}\right\Vert \\
&  :=\left\Vert \mathcal{I}^{\mu}-\mathcal{I}\right\Vert _{\diamond N}.
\end{align}
Therefore, from Eq.~(\ref{BKuniform}) we have that, for any finite energy $N$,
we may write%
\begin{equation}
\lim_{\mu\rightarrow\infty}\delta(\mu,N)=0.
\end{equation}
In other words, for any (tele-covariant) bosonic channel $\mathcal{E}$, its
teleportation simulation $\mathcal{E}^{\mu}$ converges to $\mathcal{E}$ in
energy-bounded diamond norm. The question is: \emph{Can we remove the energy
constraint?} In the next section we completely characterize the condition that
a bosonic Gaussian channel needs to satisfy in order to be simulated by
teleportation according to the uniform topology (unconstrained diamond norm).

\section{Uniform convergence in the teleportation simulation of bosonic
Gaussian channels}

Let us now consider the convergence of the teleportation simulation in the
uniform topology, i.e., according to the unconstrained diamond norm
($N\rightarrow\infty$). As we already know, this is a property that only
certain bosonic channels may have. The simplest counter-example is certainly
the identity channel for which the teleportation simulation via the
BK\ protocol strongly but not uniformly converges. See Eqs.~(\ref{strongBK})
and~(\ref{nonUNI}). As we will see below, this is also a problem for many
Gaussian channels, including all the channels that can be represented as
Gaussian unitaries, and those that can be reduced to the $B_{1}$ canonical
form via unitary transformations. The theorem below establishes the exact
condition that a single-mode Gaussian channel must have in order to be
simulated by teleportation according to the uniform topology.

\begin{theorem}
\label{theoMAIN}Consider a single-mode bosonic Gaussian channel $\mathcal{G}%
[\mathbf{T},\mathbf{N},\mathbf{d}]$ and its teleportation simulation%
\begin{equation}
\mathcal{G}^{\mu}(\rho)=\mathcal{\tilde{T}}_{aAB}\left[  \rho_{a}\otimes
(\rho_{\mathcal{G}}^{\mu})_{AB}\right]  ,
\end{equation}
where\ $\mathcal{\tilde{T}}_{aAB}$ is the LOCC$\ $of a modified BK protocol
implemented over the resource state$\ \rho_{\mathcal{G}}^{\mu}:=\mathcal{I}%
\otimes\mathcal{G}(\Phi^{\mu})$, with $\Phi^{\mu}$ being a TMSV state with
energy $\mu$. Then, we have uniform convergence%
\begin{equation}
\lim_{\mu\rightarrow\infty}\left\Vert \mathcal{G}^{\mu}-\mathcal{G}\right\Vert
_{\diamond}=0,\label{theo}%
\end{equation}
if and only if the noise matrix $\mathbf{N}$ of the Gaussian channel
$\mathcal{G}$ has full rank, i.e., $\mathrm{rank}(\mathbf{N})=2$.
\end{theorem}

\textbf{Proof}.~~Let us start by showing the implication
\begin{equation}
\mathrm{rank}(\mathbf{N})=2\Longrightarrow\text{Eq.~(\ref{theo}).}
\label{firstIM}%
\end{equation}
Consider an arbitrary single-mode Gaussian channel $\mathcal{G}[\mathbf{T}%
,\mathbf{N},\mathbf{d}]$, so that it transforms the statistical moments as in
Eq.~(\ref{mapGG}). As we know from Eq.~(\ref{BKfirst}), for any input state
$\rho_{Ra}$, we may write%
\begin{align}
\mathcal{I}_{R}\otimes\mathcal{G}^{\mu}(\rho_{Ra})  &  =\mathcal{I}_{R}%
\otimes\mathcal{G}_{B}\circ\mathcal{T}_{aAB}(\rho_{Ra}\otimes\Phi_{AB}^{\mu
})\\
&  =\mathcal{I}_{R}\otimes(\mathcal{G}_{a}\circ\mathcal{I}_{a}^{\mu}%
)(\rho_{Ra})\\
&  =\mathcal{I}_{R}\otimes\mathcal{G}_{a}^{\mu}(\rho_{Ra})
\end{align}
where $\mathcal{T}$ is the LOCC of the standard BK\ protocol and
$\mathcal{I}^{\mu}$ is the BK\ channel, which is locally equivalent to an
additive-noise Gaussian channel ($B_{2}$ form) with added noise $\xi$ as in
Eq.~(\ref{addedNOISE}). Therefore, for the Gaussian channel $\mathcal{G}^{\mu
}$ we may write the modified transformations%
\begin{equation}
\mathbf{\bar{x}\rightarrow\mathbf{T}\bar{x}}+\mathbf{d,}~~\mathbf{V}%
\rightarrow\mathbf{TVT}^{T}+\mathbf{N+}\xi\mathbf{TT}^{T}.
\end{equation}
As we can see, the transformation of the first moments is identical. By
contrast, the transformation of the second moments is characterized by the
modified noise matrix
\begin{equation}
\mathbf{N}^{\xi}=\mathbf{N+}\xi\mathbf{TT}^{T}.
\end{equation}
In order words, we may write $\mathcal{G}^{\mu}[\mathbf{\mathbf{T}}%
,\mathbf{N}^{\xi},\mathbf{d}]$.

Because $\mathcal{G}$\ and $\mathcal{G}^{\mu}$\ have the same displacement, we
can set $\mathbf{d}=\mathbf{0}$ without losing generality. Consider the
unitary reduction of $\mathcal{G}[\mathbf{T},\mathbf{N},\mathbf{0}]$ into the
corresponding canonical form $\mathcal{C}$ by means of two Gaussian unitaries
$\hat{U}_{A}$ and $\hat{U}_{B}$ as in Eq.~(\ref{unitaryEQ}). Because
$\mathbf{d}=\mathbf{0}$, we may assume that these unitaries are canonical
(i.e., with zero displacement), so that they are one-to-one with two
symplectic transformations, $\mathbf{S}_{A}$ and $\mathbf{S}_{B}$, in the
phase space. To simplify the notation define the Gaussian channels
\begin{equation}
\mathcal{U}_{A}(\rho):=\hat{U}_{A}\rho\hat{U}_{A}^{\dagger},~\mathcal{U}%
_{B}(\rho):=\hat{U}_{B}\rho\hat{U}_{B}^{\dagger}.
\end{equation}
Then we may write%
\begin{align}
\mathcal{G}  &  =\mathcal{U}_{B}\circ\mathcal{C}\circ\mathcal{U}%
_{A},\label{eq1}\\
\mathcal{G}^{\mu}  &  =\mathcal{U}_{B}\circ\mathcal{C}\circ\mathcal{U}%
_{A}\circ\mathcal{I}^{\mu}.
\end{align}
Then notice that we may re-write%
\begin{equation}
\mathcal{G}^{\mu}=\mathcal{U}_{B}\circ\mathcal{C}^{\mu}\circ\mathcal{U}_{A},
\label{eq2}%
\end{equation}
where we have defined
\begin{equation}
\mathcal{C}^{\mu}:=\mathcal{C}\circ\mathcal{U}_{A}\circ\mathcal{I}^{\mu}%
\circ\mathcal{U}_{A}^{-1}.
\end{equation}
In Appendix~\ref{APPproof} we prove the following.

\begin{lemma}
\label{LemmaMAIN}Consider a Gaussian channel $\mathcal{G}$ with $\tau
:=\det\mathbf{T}\neq1$ and $\mathrm{rank}(\mathbf{N})=2$. Then $\mathcal{C}$
and $\mathcal{C}^{\mu}$ have the same unitary dilation but different
environmental states $\rho_{e}$ and $\rho_{e}^{\mu}$, i.e., for any input
state $\rho$ we may write%
\begin{equation}
\mathcal{C}(\rho)=\mathcal{D}(\rho\otimes\rho_{e}),~~\mathcal{C}^{\mu}%
(\rho)=\mathcal{D}(\rho\otimes\rho_{e}^{\mu}),
\end{equation}
where $\mathcal{D}(\rho_{ae}):=\mathrm{Tr}_{e}\left(  \hat{U}_{ae}\rho
_{ae}\hat{U}_{ae}^{\dagger}\right)  $ with $\hat{U}_{ae}$ unitary. Furthermore%
\begin{equation}
\lim_{\mu\rightarrow\infty}F(\rho_{e}^{\mu},\rho_{e})=1.\label{FidRES}%
\end{equation}

\end{lemma}

Using this lemma in Eqs.~(\ref{eq1}) and~(\ref{eq2}) leads to%
\begin{align}
\mathcal{G}(\rho)  &  =\mathcal{U}_{B}\circ\mathcal{D}[\mathcal{U}_{A}%
(\rho)\otimes\rho_{e}],\\
\mathcal{G}^{\mu}(\rho)  &  =\mathcal{U}_{B}\circ\mathcal{D}[\mathcal{U}%
_{A}(\rho)\otimes\rho_{e}^{\mu}].
\end{align}
Clearly these relations can be extended to the presence of a reference system
$R$, so that for any input $\rho_{Ra}$, we may write%
\begin{align}
\mathcal{I}_{R}\otimes\mathcal{G}_{a}(\rho_{Ra})  &  =\mathcal{I}_{R}%
\otimes\mathcal{U}_{B}\circ\mathcal{D}[\mathcal{U}_{A}(\rho_{Ra})\otimes
\rho_{e}],\\
\mathcal{I}_{R}\otimes\mathcal{G}_{a}^{\mu}(\rho_{Ra})  &  =\mathcal{I}%
_{R}\otimes\mathcal{U}_{B}\circ\mathcal{D}[\mathcal{U}_{A}(\rho_{Ra}%
)\otimes\rho_{e}^{\mu}].
\end{align}

As a result for any $\rho_{Ra}$, we may bound the trace distance as follows%
\begin{align}
&  \left\Vert \mathcal{I}_{R}\otimes\mathcal{G}_{a}^{\mu}(\rho_{Ra}%
)-\mathcal{I}_{R}\otimes\mathcal{G}_{a}(\rho_{Ra})\right\Vert \label{data1}\\
&  =\left\Vert \mathcal{I}_{R}\otimes\mathcal{U}_{B}\circ\mathcal{D}%
[\mathcal{U}_{A}(\rho_{Ra})\otimes\rho_{e}^{\mu}]\right. \nonumber\\
&  \left.  -\mathcal{I}_{R}\otimes\mathcal{U}_{B}\circ\mathcal{D}%
[\mathcal{U}_{A}(\rho_{Ra})\otimes\rho_{e}]\right\Vert \\
&  \overset{(1)}{\leq}\left\Vert \mathcal{U}_{A}(\rho_{Ra})\otimes\rho
_{e}^{\mu}-\mathcal{U}_{A}(\rho_{Ra})\otimes\rho_{e}\right\Vert \\
&  \overset{(2)}{=}\left\Vert \rho_{e}^{\mu}-\rho_{e}\right\Vert \overset
{(3)}{\leq}2\sqrt{1-F(\rho_{e}^{\mu},\rho_{e})^{2}}, \label{data5}%
\end{align}
where we use: (1) the monotonicity under CPTP maps (including the partial
trace) (2)~multiplicity over tensor products; and (3) one of the Fuchs-van der
Graaf relations. This is a very typical computation in teleportation
stretching~\cite{PLOB} which has been adopted by several other authors in
follow-up analyses.

As we can see the upper-bound in Eq.~(\ref{data5}) does not depend on the
input state $\rho_{Ra}$. Therefore, we may extend the result to the supremum
and write%
\begin{align}
\left\Vert \mathcal{G}^{\mu}-\mathcal{G}\right\Vert _{\diamond}  &
:=\sup_{\rho_{Ra}}\left\Vert \mathcal{I}_{R}\otimes\mathcal{G}_{a}^{\mu}%
(\rho_{Ra})-\mathcal{I}_{R}\otimes\mathcal{G}_{a}(\rho_{Ra})\right\Vert
\nonumber\\
&  \leq2\sqrt{1-F(\rho_{e}^{\mu},\rho_{e})^{2}}.
\end{align}
Now, using Eq.~(\ref{FidRES}), we obtain
\begin{equation}
\lim_{\mu\rightarrow\infty}\left\Vert \mathcal{G}^{\mu}-\mathcal{G}\right\Vert
_{\diamond}=0,
\end{equation}
proving the result for $\tau:=\det\mathbf{T}\neq1$ and $\mathrm{rank}%
(\mathbf{N})=2$, i.e.,%
\begin{equation}
\left.
\begin{array}
[c]{c}%
\tau:=\det\mathbf{T}\neq1\\
\mathrm{rank}(\mathbf{N})=2~~~
\end{array}
\right\}  \Longrightarrow\text{Eq.~(\ref{theo}).}%
\end{equation}

Let us now remove the assumption $\tau:=\det\mathbf{T}\neq1$. Note that the
Gaussian channels with $\tau=1$ and $\mathrm{rank}(\mathbf{N})=2$ are those
$\mathcal{\tilde{G}}$ unitarily equivalent to the $B_{2}$ form $\mathcal{C}%
[1,2,\xi^{\prime}]$ with added noise $\xi^{\prime}\geq0$. In this case, we
dilate the form in the asymptotic single-mode representation described in
Sec.~\ref{B2section}. In other words, we may write%
\begin{align}
\mathcal{\tilde{G}}  &  =\mathcal{U}_{B}\circ\mathcal{C}[1,2,\xi^{\prime
}]\circ\mathcal{U}_{A}\\
&  =\mathcal{U}_{B}\circ\lim_{\tau\rightarrow1}\mathcal{C}[\tau,2,\bar{n}%
_{\xi^{\prime},\tau}]\circ\mathcal{U}_{A}\\
&  =\lim_{\tau\rightarrow1}~\mathcal{U}_{B}\circ\mathcal{C}[\tau,2,\bar
{n}_{\xi^{\prime},\tau}]\circ\mathcal{U}_{A}%
\end{align}
where $\bar{n}_{\xi^{\prime},\tau}:=[\xi^{\prime}(1-\tau)^{-1}-1]/2$ and it is
easy to check the commutation of the limit. Let us call $\mathcal{B}_{\tau}$
the beam-splitter dilation associated with the attenuator $C$ form
$\mathcal{C}[\tau,2,\bar{n}]$, and call $\rho_{e}(\bar{n})$ the corresponding
thermal state of the environment. Then, we may write the approximation
\begin{align}
\mathcal{\tilde{G}}  &  =\lim_{\tau\rightarrow1}~\mathcal{\tilde{G}}^{\tau
},\label{hhhh}\\
\mathcal{\tilde{G}}^{\tau}(\rho)  &  :=\mathcal{U}_{B}\circ\mathcal{B}_{\tau
}[\mathcal{U}_{A}(\rho)\otimes\rho_{e}(\bar{n}_{\xi^{\prime},\tau})].
\label{gggg}%
\end{align}

Similarly, for the teleportation-simulated channel, we may write
\begin{align}
\mathcal{\tilde{G}}^{\mu}  &  =\lim_{\tau\rightarrow1}~\mathcal{\tilde{G}%
}^{\mu,\tau},\label{hhhh2}\\
\mathcal{\tilde{G}}^{\mu,\tau}(\rho)  &  :=\mathcal{U}_{B}\circ\mathcal{B}%
_{\tau}[\mathcal{U}_{A}(\rho)\otimes\rho_{e}^{\mu}(\bar{n}_{\xi^{\prime},\tau
})], \label{gggg2}%
\end{align}
where $\rho_{e}^{\mu}(\bar{n}_{\xi^{\prime},\tau})$ is a modified
environmental state.

We can now exploit the triangle inequality. For any input $\rho$ and any
$\tau<1$, we may write
\begin{align}
\left\Vert \mathcal{\tilde{G}}^{\mu}(\rho)-\mathcal{\tilde{G}}(\rho
)\right\Vert  &  \leq\left\Vert \mathcal{\tilde{G}}^{\mu}(\rho
)-\mathcal{\tilde{G}}^{\mu,\tau}(\rho)\right\Vert \\
&  +\left\Vert \mathcal{\tilde{G}}^{\mu,\tau}(\rho)-\mathcal{\tilde{G}}^{\tau
}(\rho)\right\Vert +\left\Vert \mathcal{\tilde{G}}^{\tau}(\rho
)-\mathcal{\tilde{G}}(\rho)\right\Vert .\nonumber
\end{align}

By taking the limit for $\tau\rightarrow1$ and using Eqs.~(\ref{hhhh})
and~(\ref{hhhh2}), we find%
\begin{equation}
\left\Vert \mathcal{\tilde{G}}^{\mu}(\rho)-\mathcal{\tilde{G}}(\rho
)\right\Vert \leq\lim_{\tau\rightarrow1}\left\Vert \mathcal{\tilde{G}}%
^{\mu,\tau}(\rho)-\mathcal{\tilde{G}}^{\tau}(\rho)\right\Vert ~.
\end{equation}
Repeating previous arguments, from Eqs.~(\ref{gggg}) and~(\ref{gggg2}), we
easily derive
\begin{equation}
\left\Vert \mathcal{\tilde{G}}^{\mu,\tau}(\rho)-\mathcal{\tilde{G}}^{\tau
}(\rho)\right\Vert \leq2\sqrt{1-F[\rho_{e}^{\mu}(\bar{n}_{\xi^{\prime},\tau
}),\rho_{e}(\bar{n}_{\xi^{\prime},\tau})]^{2}},
\end{equation}
so that
\begin{equation}
\left\Vert \mathcal{\tilde{G}}^{\mu}(\rho)-\mathcal{\tilde{G}}(\rho
)\right\Vert \leq\lim_{\tau\rightarrow1}~2\sqrt{1-F[\rho_{e}^{\mu}(\bar
{n}_{\xi^{\prime},\tau}),\rho_{e}(\bar{n}_{\xi^{\prime},\tau})]^{2}}.
\end{equation}
The previous inequality holds for any input state and can be easily extended
to the presence of a reference system $R$, so that we may write
\begin{equation}
\left\Vert \mathcal{\tilde{G}}^{\mu}-\mathcal{\tilde{G}}\right\Vert
_{\diamond}\leq\lim_{\tau\rightarrow1}~2\sqrt{1-F[\rho_{e}^{\mu}(\bar{n}%
_{\xi^{\prime},\tau}),\rho_{e}(\bar{n}_{\xi^{\prime},\tau})]^{2}}.
\end{equation}
One can easily check (see Appendix~\ref{AsyAPP}), that the previous inequality
leads to uniform convergence
\begin{equation}
\lim_{\mu\rightarrow\infty}\left\Vert \mathcal{\tilde{G}}^{\mu}%
-\mathcal{\tilde{G}}\right\Vert _{\diamond}=0~,
\end{equation}
completing the proof of the implication in Eq.~(\ref{firstIM}).

Let us now show the opposite implication
\begin{equation}
\mathrm{rank}(\mathbf{N})=2\Longleftarrow\text{Eq.~(\ref{theo}),}%
\end{equation}
or, equivalently,
\begin{equation}
\mathrm{rank}(\mathbf{N})<2\Longrightarrow\text{No uniform convergence.}%
\end{equation}
Note that Gaussian channels with $\mathrm{rank}(\mathbf{N})<2$ are the
identity channel $B_{2}(Id)$, having zero rank,\ and the $B_{1}$ form, having
unit rank. We already know that there is no uniform convergence in the
teleportation simulation of the identity channel and this property trivially
extends to the teleportation simulation $\mathcal{U}^{\mu}=\mathcal{U}%
\circ\mathcal{I}^{\mu}$ of any Gaussian unitary $\mathcal{U}$. In fact, it is
easy to check that%
\begin{equation}
\left\Vert \mathcal{U}^{\mu}-\mathcal{U}\right\Vert _{\diamond}=\left\Vert
\mathcal{I}^{\mu}-\mathcal{I}\right\Vert _{\diamond}=2~,
\end{equation}
due to invariance under unitaries. For the $B_{1}$ form $\mathcal{\tilde{C}%
}=\mathcal{C}[1,1,0]$, we now explicitly show that there is no uniform
convergence in its teleportation simulation. Let us consider the simulation
$\mathcal{\tilde{C}}^{\mu}$ by means of a $\mu$-energy BK protocol and
consider an input TMSV state $\Phi_{Ra}^{\tilde{\mu}}$ with diverging energy
$\tilde{\mu}$. We have the two output states%
\begin{equation}
\rho_{Ra}^{\tilde{\mu}}:=\mathcal{I}_{R}\otimes\mathcal{\tilde{C}}_{a}%
(\Phi_{Ra}^{\tilde{\mu}}),~\rho_{Ra}^{\mu,\tilde{\mu}}:=\mathcal{I}_{R}%
\otimes\mathcal{\tilde{C}}_{a}^{\mu}(\Phi_{Ra}^{\tilde{\mu}}).
\end{equation}
In particular, note that $\rho_{Ra}^{\mu,\tilde{\mu}}$ is a Gaussian state
with CM%
\begin{equation}
\mathbf{V}^{\mu,\tilde{\mu}}=\left(
\begin{array}
[c]{cccc}%
\tilde{\mu} & 0 & \sqrt{\tilde{\mu}^{2}-1} & 0\\
0 & \tilde{\mu} & 0 & -\sqrt{\tilde{\mu}^{2}-1}\\
\sqrt{\tilde{\mu}^{2}-1} & 0 & \tilde{\mu}+\xi & 0\\
0 & -\sqrt{\tilde{\mu}^{2}-1} & 0 & \tilde{\mu}+\xi+1
\end{array}
\right)  ,
\end{equation}
where $\xi$ is the added noise associated with the BK\ protocol and depends on
$\mu$ according to Eq.~(\ref{addedNOISE}). Using Eq.~(\ref{FuchsGraaf}) we may
write%
\begin{equation}
\left\Vert \rho_{Ra}^{\mu,\tilde{\mu}}-\rho_{Ra}^{\tilde{\mu}}\right\Vert
\geq2\left[  1-F\left(  \rho_{Ra}^{\mu,\tilde{\mu}},\rho_{Ra}^{\tilde{\mu}%
}\right)  \right]  .
\end{equation}
Then, by computing the fidelity~\cite{banchiPRL2015} and expanding in
$\tilde{\mu}$, we obtain%
\begin{equation}
F\left(  \rho_{Ra}^{\mu,\tilde{\mu}},\rho_{Ra}^{\tilde{\mu}}\right)  \simeq
O(\tilde{\mu}^{-1/4}),
\end{equation}
so that
\begin{equation}
\lim_{\tilde{\mu}\rightarrow\infty}\left\Vert \rho_{Ra}^{\mu,\tilde{\mu}}%
-\rho_{Ra}^{\tilde{\mu}}\right\Vert =2,
\end{equation}
which clearly implies $\left\Vert \mathcal{\tilde{C}}^{\mu}-\mathcal{\tilde
{C}}\right\Vert _{\diamond}=2$. Then, we may extend the result to any Gaussian
channel which is unitarily equivalent to the $B_{1}$ form. Consider
Eqs.~(\ref{eq1}) and~(\ref{eq2}) with $\mathcal{\tilde{C}}=\mathcal{C}%
[1,1,0]$, i.e,%
\begin{equation}
\mathcal{G}=\mathcal{U}_{B}\circ\mathcal{\tilde{C}}\circ\mathcal{U}%
_{A},~~\mathcal{G}^{\mu}=\mathcal{U}_{B}\circ\mathcal{\tilde{C}}^{\mu}%
\circ\mathcal{U}_{A},
\end{equation}
where
\begin{equation}
\mathcal{\tilde{C}}^{\mu}:=\mathcal{\tilde{C}}\circ\mathcal{U}_{A}%
\circ\mathcal{I}^{\mu}\circ\mathcal{U}_{A}^{-1}. \label{fromrr}%
\end{equation}
Assume the input state $\Psi_{Ra}^{\tilde{\mu}}:=\mathcal{I}_{R}%
\otimes\mathcal{U}_{A}^{-1}(\Phi_{Ra}^{\tilde{\mu}})$, so that we have the two
output states%
\begin{align}
\rho_{Ra}^{\tilde{\mu}}  &  :=\mathcal{I}_{R}\otimes\mathcal{G}_{a}(\Psi
_{Ra}^{\tilde{\mu}})=\mathcal{I}_{R}\otimes\mathcal{U}_{B}\circ\mathcal{\tilde
{C}}(\Phi_{Ra}^{\tilde{\mu}}),\\
\rho_{Ra}^{\mu,\tilde{\mu}}  &  :=\mathcal{I}_{R}\otimes\mathcal{G}_{a}^{\mu
}(\Psi_{Ra}^{\tilde{\mu}})=\mathcal{I}_{R}\otimes\mathcal{U}_{B}%
\circ\mathcal{\tilde{C}}^{\mu}(\Phi_{Ra}^{\tilde{\mu}}).
\end{align}
Because the fidelity is invariant under unitaries, we may neglect
$\mathcal{U}_{B}$ and write%
\begin{equation}
F\left(  \rho_{Ra}^{\mu,\tilde{\mu}},\rho_{Ra}^{\tilde{\mu}}\right)  =F\left[
\mathcal{I}_{R}\otimes\mathcal{\tilde{C}}^{\mu}(\Phi_{Ra}^{\tilde{\mu}%
}),\mathcal{I}_{R}\otimes\mathcal{\tilde{C}}(\Phi_{Ra}^{\tilde{\mu}})\right]
.
\end{equation}
Let us derive the CM $\mathbf{\tilde{V}}^{\mu,\tilde{\mu}}$ of the state
$\mathcal{I}_{R}\otimes\mathcal{\tilde{C}}^{\mu}(\Phi_{Ra}^{\tilde{\mu}})$.
Starting from the CM $\mathbf{V}^{\mu}$ of the TMSV in Eq.~(\ref{TMSVstate})
and applying Eq.~(\ref{fromrr}), we easily see that this CM is given by%
\begin{equation}
\mathbf{\tilde{V}}^{\mu,\tilde{\mu}}=\mathbf{V}^{\mu}+\mathbf{0}\oplus\left[
\xi\mathbf{S}_{A}\mathbf{S}_{A}^{T}+\text{\textrm{diag}}(0,1)\right]  ,
\end{equation}
where $\mathbf{0}$ is the $2\times2$ zero matrix, and $\mathbf{S}_{A}$\ is the
symplectic matrix associated with the Gaussian unitary $\mathcal{U}_{A}$
(which can be taken to be canonical without losing generality). Let us set
\begin{equation}
\mathbf{S}_{A}=\left(
\begin{array}
[c]{cc}%
a & c\\
d & b
\end{array}
\right)  ,
\end{equation}
where the elements are real values such that $\det\mathbf{S}_{A}=+1$ (because
$\mathbf{S}_{A}$ is symplectic). Then, we may compute the fidelity and expand
it at the leading order in $\tilde{\mu}$, finding%
\begin{align}
&  F\left[  \mathcal{I}_{R}\otimes\mathcal{\tilde{C}}^{\mu}(\Phi_{Ra}%
^{\tilde{\mu}}),\mathcal{I}_{R}\otimes\mathcal{\tilde{C}}(\Phi_{Ra}%
^{\tilde{\mu}})\right]  ^{4}\nonumber\\
&  \simeq\gamma\tilde{\mu}^{-1}+O(\tilde{\mu}^{-3/2}),\\
\gamma &  :=\frac{a^{2}+c^{2}+2\xi}{2\xi(a^{2}+c^{2}+\xi)^{2}}>0.
\end{align}
Clearly, this implies $\left\Vert \mathcal{G}^{\mu}-\mathcal{G}\right\Vert
_{\diamond}=2$ for any Gaussian channel unitarily equivalent to the $B_{1}$
form.$~\blacksquare$

\bigskip

Note that the rank of the noise matrix $\mathbf{N}$ is indeed a fundamental
quantity in the previous proof. Given a single-mode Gaussian channel
$\mathcal{G}[\mathbf{T},\mathbf{N},\mathbf{d}]$, consider its teleportation
simulation $\mathcal{G}^{\mu}[\mathbf{\mathbf{T}},\mathbf{N}^{\xi}%
,\mathbf{d}]$. For all channels with $\mathrm{rank}(\mathbf{N})=2$, we may
write%
\begin{equation}
\mathrm{rank}(\mathbf{N}^{\xi})=\mathrm{rank}(\mathbf{N})~~\text{for any }%
\xi\text{.}%
\end{equation}
This means that $\mathcal{G}^{\mu}$ may have the same canonical form and,
therefore, the same unitary dilation as $\mathcal{G}$. By contrast, for
Gaussian channels with $\mathrm{rank}(\mathbf{N})<2$, such as the identity
channel or the $B_{1}$ form, we can see that we have $\mathrm{rank}%
(\mathbf{N}^{\xi})>\mathrm{rank}(\mathbf{N})$ for $\xi\neq0$, so that the
canonical form changes its class because of the teleportation simulation. As a
result, the dilation changes and the data-processing bound in
Eqs.~(\ref{data1})-(\ref{data5}) cannot be applied.

\begin{remark}
Our Theorem~\ref{theoMAIN} straightforwardly solidifies all the
claims of uniform convergence discussed in
Ref.~\cite[v4]{WildePLAG} for very specific channels. Note that
Ref.~\cite[v4]{WildePLAG} did not consider an arbitrary
single-mode Gaussian channel, but only the canonical forms $C$ and
$B_{2}$, without considering the action of input-output Gaussian
unitaries. Furthermore, the other canonical forms, together with
the Gaussian channels unitarily equivalent to them, were also not
considered in Ref.~\cite[v4]{WildePLAG}.
\end{remark}

\begin{remark}
After our Theorem~\ref{theoMAIN} was publicly available on the
arXiv~\cite{nostro}, we noticed that Ref.~\cite{WildePLAG} was
later updated into its $5^{th}$ version, where our key observation
on the rank of the noise matrix of the Gaussian channel has been
inserted (with no discussion of Ref.~\cite{nostro}). In fact, see
Theorem 6 in Ref.~\cite[v5]{WildePLAG} which appears to be an
immediate extension of our Theorem~\ref{theoMAIN}.
\end{remark}

\section{Teleportation simulation of bosonic channels in adaptive
protocols\label{teleSEC}}

We now discuss the teleportation simulation of bosonic channels within the
context of adaptive protocols. This treatment is particularly important for
its implications in quantum and private communications.

\subsection{Adaptive protocols}

In an adaptive protocol, Alice and Bob, are connected by a quantum channel
$\mathcal{E}$ at the ends of which they apply the most general quantum
operations (QOs) allowed by quantum mechanics. If the task of the protocol is
quantum channel discrimination or estimation, then Alice and Bob are the same
entity~\cite{ReviewMETRO,PirCo}. However, if the task of the protocol is
quantum/private communication, Alice and Bob are distinct remote users and
their QOs consist of local operations (LOs) assisted by unlimited and two-way
classical communication (CC), briefly called adaptive LOCCs~\cite{PLOB,TQC}.
For simplicity, we consider here the second case only, i.e., communication.

The adaptive LOCCs are interleaved with the various transmissions through the
channel. A compact formulation of the adaptive communication protocol goes as
follows. Alice and Bob have local registers $\mathbf{a}$ and $\mathbf{b}$
prepared in some fundamental state $\rho_{\mathbf{a}}^{0}\otimes
\rho_{\mathbf{b}}^{0}$. They apply a first adaptive LOCC $\Lambda_{0}$ so that
$\rho_{\mathbf{ab}}^{0}=\Lambda_{0}(\rho_{\mathbf{a}}^{0}\otimes
\rho_{\mathbf{b}}^{0})$. Then, Alice transmits one of her modes $a_{1}%
\in\mathbf{a}$ through the channel $\mathcal{E}$. Bob gets a corresponding
output mode $b_{1}$ which is included in his register $b_{1}\mathbf{b}%
\rightarrow\mathbf{b}$. The two parties apply another adaptive LOCC
$\Lambda_{1}$ to their updated registers, before the second transmission
through the channel, and so on. After $n$ uses of the channel we have the
output state
\begin{equation}
\rho_{\mathbf{ab}}^{n}=\Lambda_{n}\circ\mathcal{E}\circ\Lambda_{n-1}%
\cdots\circ\Lambda_{1}\circ\mathcal{E}\circ\Lambda_{0}(\rho_{\mathbf{a}}%
^{0}\otimes\rho_{\mathbf{b}}^{0}), \label{ada1}%
\end{equation}
where we assume that channel $\mathcal{E}$ is applied to the input system
$a_{i}$ in the $i$-th transmission, i.e., $\mathcal{E}=\mathcal{I}%
_{\mathbf{a}}\otimes\mathcal{E}_{a_{i}}\otimes\mathcal{I}_{\mathbf{b}}$.

Assume that Alice and Bob generate an output state $\rho_{\mathbf{ab}}^{n}$
which is epsilon-close to a private state~\cite{KD} $\phi_{n}$ with $nR_{n}$
secret bits, i.e., we have the trace-norm inequality
\begin{equation}
\left\Vert \rho_{\mathbf{ab}}^{n}-\phi_{n}\right\Vert \leq\varepsilon~.
\end{equation}
Then, we say that the sequence $\mathcal{P}=\{\Lambda_{0},\ldots,\Lambda
_{n}\}$ represents an $(n,\varepsilon,R_{n}^{\varepsilon})$ adaptive key
generation protocol. By optimizing over all the protocols, we may write
\begin{equation}
K(\mathcal{E},n,\varepsilon)=\sup_{\mathcal{P}}R_{n}^{\varepsilon}~.
\end{equation}
Taking the limit for large $n$ and small $\varepsilon$, one gets the secret
key capacity of the channel $K(\mathcal{E})$.

\subsection{Simulation and \textquotedblleft peeling\textquotedblright\ of
adaptive protocols}

Consider a tele-covariant bosonic channel $\mathcal{E}$. Then, let us assume a
finite-energy BK\ protocol $(\mathcal{T},\Phi^{\mu})$ so that the bosonic
channel $\mathcal{E}$\ is approximated by its teleportation simulation
$\mathcal{E}^{\mu}=\mathcal{E}\circ\mathcal{I}^{\mu}$, where $\mathcal{I}%
^{\mu}$ is the usual BK channel. In the adaptive protocol, we may then replace
each instance of $\mathcal{E}$ with $\mathcal{E}^{\mu}$. This leads to a
simulated protocol, with simulated output state $\rho_{\mathbf{ab}}^{\mu,n}$.
A crucial step is to show how the error in the channel simulation
$\mathcal{E}^{\mu}\simeq\mathcal{E}$ propagates to the output state
$\rho_{\mathbf{ab}}^{\mu,n}\simeq\rho_{\mathbf{ab}}^{n}$ after $n$ uses of the
adaptive protocol. This is done by adopting a peeling technique which suitably
exploits data processing and the triangle inequality.

After $n$ uses of the simulated channel, we may write the output state as
\begin{equation}
\rho_{\mathbf{ab}}^{\mu,n}=\Lambda_{n}\circ\mathcal{E}^{\mu}\circ\Lambda
_{n-1}\cdots\circ\Lambda_{1}\circ\mathcal{E}^{\mu}\circ\Lambda_{0}%
(\rho_{\mathbf{a}}^{0}\otimes\rho_{\mathbf{b}}^{0}), \label{ada22}%
\end{equation}
where we assume that channel $\mathcal{E}^{\mu}$ is applied to the input
system $a_{i}$ in the $i$-th transmission, i.e., $\mathcal{E}^{\mu
}=\mathcal{I}_{\mathbf{a}}\otimes\mathcal{E}_{a_{i}}^{\mu}\otimes
\mathcal{I}_{\mathbf{b}}$. We now want to evaluate the trace distance
$\left\Vert \rho_{\mathbf{ab}}^{\mu,n}-\rho_{\mathbf{ab}}^{n}\right\Vert $ and
show that this can be suitably bounded for any $n$. Let us start with the most
elegant and rigorous approach, which has been discussed in Ref.~\cite{TQC} and
directly comes from techniques in PLOB~\cite{PLOB}.

\subsubsection{Peeling in the bounded-uniform topology\label{peelingBB}}

The most rigorous way to show the peeling procedure is by using the
energy-constrained diamond distance defined in Eq.~(\ref{boundedDIAMOND}) for
an arbitrary but finite energy constraint $N$. As we have already written in
Eq.~(\ref{already}), given an input alphabet $\mathcal{D}_{N}$ with maximum
energy $N$, the simulation error between a bosonic channel $\mathcal{E}$ and
its simulation $\mathcal{E}^{\mu}$ via the $\mu$-energy Braunstein-Kimble
protocol can be written as~\cite{PLOB,TQC}
\begin{equation}
\delta(\mu,N):=\left\Vert \mathcal{E}^{\mu}-\mathcal{E}\right\Vert _{\diamond
N}\leq\left\Vert \mathcal{I}^{\mu}-\mathcal{I}\right\Vert _{\diamond N}~.
\end{equation}
For any finite value of the constraint $N$, we may take the limit in
$\mu\rightarrow\infty$ and write $\delta(\mu,N)\rightarrow0$, thanks to the
bounded-uniform convergence $\left\Vert \mathcal{I}^{\mu}-\mathcal{I}%
\right\Vert _{\diamond N}\overset{\mu\rightarrow\infty}{\rightarrow}0$.

Let us now express the output error $\left\Vert \rho_{\mathbf{ab}}^{\mu
,n}-\rho_{\mathbf{ab}}^{n}\right\Vert $ in terms of the channel error
$\delta(\mu,N)$. For simplicity, let us start by assuming $n=2$. From
Eqs.~(\ref{ada1}) and~(\ref{ada22}) we may then write the peeling
as~\cite{PLOB}
\begin{align}
&  \Vert\rho_{\mathbf{ab}}^{\mu,2}-\rho_{\mathbf{ab}}^{2}\Vert\nonumber\\
&  \overset{(1)}{\leq}\Vert\mathcal{E}^{\mu}\circ\Lambda_{1}\circ
\mathcal{E}^{\mu}(\rho_{\mathbf{ab}}^{0})-\mathcal{E}\circ\Lambda_{1}%
\circ\mathcal{E}(\rho_{\mathbf{ab}}^{0})\Vert\label{peel11}\\
&  \overset{(2)}{\leq}\Vert\mathcal{E}^{\mu}\circ\Lambda_{1}\circ
\mathcal{E}^{\mu}(\rho_{\mathbf{ab}}^{0})-\mathcal{E}\circ\Lambda_{1}%
\circ\mathcal{E}^{\mu}(\rho_{\mathbf{ab}}^{0})\Vert\nonumber\\
&  +\Vert\mathcal{E}\circ\Lambda_{1}\circ\mathcal{E}^{\mu}(\rho_{\mathbf{ab}%
}^{0})-\mathcal{E}\circ\Lambda_{1}\circ\mathcal{E}(\rho_{\mathbf{ab}}%
^{0})\Vert\\
&  \overset{(1)}{\leq}\Vert\mathcal{E}^{\mu}(\rho_{\mathbf{ab}}^{0}%
)-\mathcal{E}(\rho_{\mathbf{ab}}^{0})\Vert\nonumber\\
&  +\Vert\mathcal{E}^{\mu}[\Lambda_{1}\circ\mathcal{E}^{\mu}(\rho
_{\mathbf{ab}}^{0})]-\mathcal{E}[\Lambda_{1}\circ\mathcal{E}^{\mu}%
(\rho_{\mathbf{ab}}^{0})]\Vert\label{joe}\\
&  \overset{(3)}{\leq}2\Vert\mathcal{E}^{\mu}-\mathcal{E}\Vert_{\Diamond
N}=2\delta(\mu,N), \label{casen2}%
\end{align}
where we use: (1)~The monotonicity of the trace distance under CPTP maps; (2)
the triangle inequality; and (3)~the energy-constrained diamond distance.
Generalization to $n\geq2$ gives the desired result
\begin{equation}
\left\Vert \rho_{\mathbf{ab}}^{\mu,n}-\rho_{\mathbf{ab}}^{n}\right\Vert \leq
n\delta(\mu,N). \label{OOO1}%
\end{equation}
Now, for any finite $N$, we may take the limit for large $\mu$ and write
\begin{equation}
\left\Vert \rho_{\mathbf{ab}}^{\mu,n}-\rho_{\mathbf{ab}}^{n}\right\Vert
\rightarrow0.
\end{equation}

\subsubsection{Peeling in the uniform topology}

Consider an adaptive protocol over a bosonic Gaussian channel $\mathcal{G}%
[\mathbf{T},\mathbf{N},\mathbf{d}]$ with $\mathrm{rank}(N)=2$. In this case,
we now know that we can remove the energy constraint in the diamond distance
and write the following uniform convergence result
\begin{equation}
\left\Vert \mathcal{G}^{\mu}-\mathcal{G}\right\Vert _{\diamond}\overset
{\mu\rightarrow\infty}{\rightarrow}0~,
\end{equation}
where $\mathcal{G}^{\mu}$\ is the teleportation simulation of $\mathcal{G}%
$.\ It is clear that we can repeat the peeling and write
\begin{equation}
\left\Vert \rho_{\mathbf{ab}}^{\mu,n}-\rho_{\mathbf{ab}}^{n}\right\Vert \leq
n\left\Vert \mathcal{G}^{\mu}-\mathcal{G}\right\Vert _{\diamond}\overset
{\mu\rightarrow\infty}{\rightarrow}0. \label{OOO2}%
\end{equation}

\subsubsection{Peeling in the strong topology\label{peelingSS}}

The procedure can be trivially modified for strong convergence. In fact,
starting from Eq.~(\ref{joe}) we may write
\begin{align}
&  \Vert\rho_{\mathbf{ab}}^{\mu,2}-\rho_{\mathbf{ab}}^{2}\Vert\nonumber\\
&  \leq\Vert\mathcal{E}^{\mu}(\rho_{\mathbf{ab}}^{0})-\mathcal{E}%
(\rho_{\mathbf{ab}}^{0})\Vert\nonumber\\
&  +\Vert\mathcal{E}^{\mu}[\Lambda_{1}\circ\mathcal{E}^{\mu}(\rho
_{\mathbf{ab}}^{0})]-\mathcal{E}[\Lambda_{1}\circ\mathcal{E}^{\mu}%
(\rho_{\mathbf{ab}}^{0})]\Vert\\
&  =\Vert\mathcal{E}^{\mu}(\rho_{\mathbf{ab}}^{0})-\mathcal{E}(\rho
_{\mathbf{ab}}^{0})\Vert+\Vert\mathcal{E}^{\mu}(\rho_{\mathbf{ab}}^{\mu
,1})-\mathcal{E}(\rho_{\mathbf{ab}}^{\mu,1})\Vert,
\end{align}
where $\rho_{\mathbf{ab}}^{\mu,1}:=\Lambda_{1}\circ\mathcal{E}^{\mu}%
(\rho_{\mathbf{ab}}^{0})$ is an energy-constrained state. Then, we may write%
\begin{equation}
\Vert\rho_{\mathbf{ab}}^{\mu,2}-\rho_{\mathbf{ab}}^{2}\Vert\leq2\sup
_{\rho_{\mathbf{ab}}}\Vert\mathcal{E}^{\mu}(\rho_{\mathbf{ab}})-\mathcal{E}%
(\rho_{\mathbf{ab}})\Vert.
\end{equation}
Similarly, for $n$ uses, one derives%
\begin{equation}
\Vert\rho_{\mathbf{ab}}^{\mu,n}-\rho_{\mathbf{ab}}^{n}\Vert\leq n\sup
_{\rho_{\mathbf{ab}}}\Vert\mathcal{E}^{\mu}(\rho_{\mathbf{ab}})-\mathcal{E}%
(\rho_{\mathbf{ab}})\Vert. \label{OOO3}%
\end{equation}
Now using the strong convergence of the BK\ protocol, one gets%
\begin{equation}
\Vert\rho_{\mathbf{ab}}^{\mu,n}-\rho_{\mathbf{ab}}^{n}\Vert\overset
{\mu\rightarrow\infty}{\rightarrow}0.
\end{equation}
This type of peeling is a trivial modification of the one presented in
Sec.~\ref{peelingBB} and already adopted in PLOB and Ref.~\cite{TQC}.

\begin{remark}
Contrary to what claimed in the various arXiv versions of~Ref.~\cite[v1-v5]%
{WildePLAG}, none of the peeling techniques presented in this section can
actually be found in WTB~\cite{WildeFollowup}. Therefore, the arguments
presented in Ref.~\cite[v1-v5]{WildePLAG} are, as a matter of fact, a direct
confirmation of the technical gaps and issues of WTB~\cite{WildeFollowup}.
\end{remark}

\section{Implications for quantum and private communications}

Here we discuss how the previous notions can be used to rigorously prove the
claims presented in WTB on the strong converse bounds for private
communication over Gaussian channels. In order to clarify the technical
problems, we first provide some preliminary notions, starting from the weak
converse bounds established in PLOB and how the follow-up WTB attempted to
show their strong converse property. Then, we discuss the basic technical
errors in WTB and how these can be fixed by adopting a rigorous treatment of
the BK\ protocol in adaptive protocols. Our proofs expand the very first one
provided in Ref.~\cite{TQC} and based on the bounded-uniform convergence of
the BK\ protocol.

\subsection{Background}

Quantum and private communications over optical channels are inevitably
limited by the presence of loss. In fact, the maximum number of secret bits
that can be distributed over an optical fiber or a free-space link cannot be
arbitrary but scales as $\simeq\tau$, where $\tau$ is the transmissivity of
the communication channel. This is a fundamental rate-loss law that has
attracted a lot of attention in the past years~\cite{Rev1,Rev2,TGW,PLOB}. In
2009, Pirandola-Patr\'{o}n-Braunstein-Lloyd~\cite{Rev2} used the reverse
coherent information~\cite{Rev1} to compute the best-known achievable rate of
$-\log_{2}(1-\tau)$ secret bits per use.

Later, in 2014, Takeoka-Guha-Wilde~\cite{TGW} computed the first upper bound
$\log_{2}[(1+\tau)(1-\tau)]$ by resorting to the squashed
entanglement~\cite{Squash}. Finally, in 2015, PLOB~\cite{PLOB} exploited
quantum teleportation~\cite{Tele1,Tele2,telereview} and the relative entropy
of entanglement~\cite{REE1,REE2,REE3} to establish $-\log_{2}(1-\tau)$ as an
upper bound, therefore discovering the secret-key capacity $K$ of the
pure-loss channel. This result is also known as the PLOB bound. It was
promptly generalized to repeater-assisted lossy communications~\cite{netPAPER}
and its strong converse property was later investigated by
WTB~\cite{WildeFollowup}.

In the bosonic setting, PLOB~\cite{PLOB} proved weak converse upper bounds for
the private communication over single-mode phase-insensitive Gaussian
channels~\cite{RMP}. In fact, let us introduce the entropic function
\begin{equation}
h(x):=(x+1)\log_{2}(x+1)-x\log_{2}x. \label{hEntropyMAIN}%
\end{equation}
Then, for a thermal-loss channel $\mathcal{C}_{\tau,\bar{n}}^{\text{loss}%
}=\mathcal{C}[\tau,2,\bar{n}]$ with transmissivity $\tau\in\lbrack0,1]$ and
mean thermal number $\bar{n}$ (canonical form $C$), one has~\cite{PLOB}%
\begin{align}
K(\mathcal{C}_{\tau,\bar{n}}^{\text{loss}})  &  \leq\Phi(\mathcal{C}%
_{\tau,\bar{n}}^{\text{loss}})\label{LossUB}\\
&  :=\left\{
\begin{array}
[c]{l}%
-\log_{2}\left[  (1-\tau)\tau^{\bar{n}}\right]  -h(\bar{n}),~~\text{for~}%
\bar{n}<\frac{\tau}{1-\tau},\\
0\text{,~~otherwise.}%
\end{array}
\right. \nonumber
\end{align}
For a quantum amplifier $\mathcal{C}_{\tau,\bar{n}}^{\text{amp}}%
=\mathcal{C}[\tau,2,\bar{n}]$ with gain $\tau>1$ and mean thermal number
$\bar{n}$ (canonical form $C$), one has the following bound~\cite{PLOB}%
\begin{align}
K(\mathcal{C}_{\tau,\bar{n}}^{\text{amp}})  &  \leq\Phi(\mathcal{C}_{\tau
,\bar{n}}^{\text{amp}})\label{ampliUB}\\
&  :=\left\{
\begin{array}
[c]{l}%
\log_{2}\left(  \dfrac{\tau^{\bar{n}+1}}{\tau-1}\right)  -h(\bar
{n}),~~\text{for~}\bar{n}<(\tau-1)^{-1},\\
0\text{,~~otherwise.}%
\end{array}
\right. \nonumber
\end{align}
Finally, for an additive-noise Gaussian channel $\mathcal{C}_{\xi}%
^{\text{add}}=\mathcal{C}[1,2,\xi]$ with added noise $\xi\geq0$ (canonical
form $B_{2}$), we may write~\cite{PLOB}%
\begin{equation}
K(\mathcal{C}_{\xi}^{\text{add}})\leq\Phi(\mathcal{C}_{\xi}^{\text{add}%
}):=\left\{
\begin{array}
[c]{l}%
\frac{\xi-1}{\ln2}-\log_{2}\xi,~~\text{for~}\xi<1,\\
0\text{,~~otherwise.}%
\end{array}
\right.  \label{AdditiveUB}%
\end{equation}

\begin{remark}
In a talk~\cite{WildeQcrypt}, an author wrongly claimed that the bounds in
Eqs.~(\ref{LossUB}), (\ref{ampliUB}) and~(\ref{AdditiveUB}) would explode due
to a technical issue related with the unboundedness of the \textquotedblleft
shield size\textquotedblright\ of the continuous-variable private state. This
is not the case because the dimension of the private state was suitably
\textbf{truncated} already in the first 2015 proof given by PLOB. See
Ref.~\cite[Sec. III]{TQC} for further discussions and more details
demystifying these claims.
\end{remark}

\begin{remark}
As a direct result of his basic misunderstanding of the 2015 proof
given by PLOB (see previous remark), the same author started to
(unfairly) credit his follow-up work WTB~\cite{WildeFollowup} for
the proof of the weak-converse upper bounds in
Eqs.~(\ref{LossUB}), (\ref{ampliUB}) and~(\ref{AdditiveUB}). As
one can easily check on the public arXiv, these bounds were fully
established by PLOB in 2015, several months before WTB even made
its first appearance (29 Feb 2016). From the chronology on the
arXiv, one can also easily check that the main tools used in WTB
were directly taken from PLOB. In this context, Ref.~\cite{TQC}
fully clarifies how WTB is a direct follow-up work which is
heavily based on results and tools in PLOB. This aspect is also
very clear from the first arXiv version of WTB, where the
presentation of previous results was sufficiently fair, but then
its terminology was suddenly changed in its published version,
where unfair claims have been made.
\end{remark}

\subsection{General problems with the strong converse bounds claimed in WTB}

Several months after the first version of PLOB, the follow-up paper
WTB~\cite{WildeFollowup} also appeared on the arXiv. One of the main aims of
WTB was to show that PLOB's weak converse bounds for single-mode
phase-insensitive bosonic Gaussian channels in Eqs.~(\ref{LossUB}%
)-(\ref{AdditiveUB}) also have the strong converse property. Recall that a
weak converse bound means that \textit{perfect} secret keys cannot be
established at rates exceeding the bound. A strong converse bound is a
refinement according to which even \textit{imperfect} secret keys
($\varepsilon$-secure with $\varepsilon>0$) cannot be generated above the
bound. In terms of methodology, WTB widely exploited the tools previously
introduced by PLOB, in particular, the notion of a channel's REE and the
adaptive-to-block simplification via teleportation stretching. The combination
of these two ingredients allowed PLOB (and later WTB) to write single-letter
bounds in terms of the REE. However, differently from PLOB, WTB did not
explicitly prove its statements for two main reasons:

\begin{description}
\item[(1)] WTB did not show how the error affecting the simulation of the
bosonic channels is propagated to the output state of an adaptive protocol;

\item[(2)] WTB did not show that such error converges to zero.
\end{description}

\noindent As a result of these two points, the bounds in WTB were not shown
for adaptive protocols and, as presented there, they were technically equal to
infinity. Let us describe these issues in details in the following section.

\subsection{Strong converse claims}

In \cite[Theorem~24]{WildeFollowup}, WTB\ made the following claims on the
strong-converse bound for single-mode phase-insensitive Gaussian channels.

\bigskip

\noindent\textbf{WTB claims~(\cite{WildeFollowup}).}~Consider an $\varepsilon
$-secure key generation protocol over $n$ uses of a phase-insensitive
canonical form $\mathcal{C}$, which may be a thermal-loss channel
($\mathcal{C}_{\tau,\bar{n}}^{\text{loss}}$), a quantum amplifier
($\mathcal{C}_{\tau,\bar{n}}^{\text{amp}}$) or an additive-noise Gaussian
channel ($\mathcal{C}_{\xi}^{\text{add}}$). For any $\varepsilon\in(0,1)$ and
$n\geq1$, one may write the following upper bound for the secret key rate%
\begin{equation}
K(\mathcal{C})\leq\Phi(\mathcal{C})+\sqrt{\frac{V(\mathcal{C})}%
{n(1-\varepsilon)}}+\frac{C(\varepsilon)}{n}~, \label{strongconv}%
\end{equation}
where $\Phi(\mathcal{C})$ is PLOB's weak converse bound given in
Eqs.~(\ref{LossUB})-(\ref{AdditiveUB}), $V(\mathcal{C})$ is a suitable
\textquotedblleft unconstrained relative entropy variance\textquotedblright,
and%
\begin{equation}
C(\varepsilon):=\log_{2}6+2\log_{2}\left(  \frac{1+\varepsilon}{1-\varepsilon
}\right)  .
\end{equation}
In particular, for a pure loss channel ($\mathcal{C}_{\tau,0}^{\text{loss}}$)
and a quantum-limited amplifier ($\mathcal{C}_{\tau,0}^{\text{amp}}$), one
would have%
\begin{equation}
K(\mathcal{C})\leq\Phi(\mathcal{C})+\frac{C(\varepsilon)}{n}~. \label{cc11}%
\end{equation}

The above claims are obtained starting from a teleportation simulation based
on the BK protocol with finite energy $\mu$ and then taking the limit of
$\mu\rightarrow\infty$ (following PLOB). For any security parameter
$\varepsilon\in(0,1)$, number of channel uses $n\geq1$ and simulation energy
$\mu$ with \textquotedblleft infidelity\textquotedblright\ $\varepsilon
_{\text{TP}}(n,\mu)$, one may write the following upper bound for the secret
key rate of a phase insensitive canonical form $\mathcal{C}$%
\begin{equation}
K(\mathcal{C})\leq\Phi(\mathcal{C})+\Delta(n,\mu). \label{followPLOB}%
\end{equation}
At fixed $n$ and large $\mu$, $\Delta(n,\mu)$ has the expansion%
\begin{equation}
\Delta(n,\mu)\simeq\sqrt{\frac{V(\mathcal{C})+O(\mu^{-1})}{n[1-\varepsilon
(n,\mu)]}}+\frac{C[\varepsilon(n,\mu)]}{n}+O(\mu^{-1})~, \label{eqrr}%
\end{equation}
where $\varepsilon(n,\mu)$ is an overall error defined as%
\begin{equation}
\varepsilon(n,\mu):=\min\left\{  1,\left[  \sqrt{\varepsilon}+\sqrt
{\varepsilon_{\text{TP}}(n,\mu)}\right]  ^{2}\right\}  , \label{errorTT}%
\end{equation}
where $\varepsilon_{\text{TP}}$ is associated with the teleportation
simulation. For a pure loss channel ($\mathcal{C}_{\tau,0}^{\text{loss}}$) and
a quantum-limited amplifier ($\mathcal{C}_{\tau,0}^{\text{amp}}$), one has
Eq.~(\ref{followPLOB}), with
\begin{equation}
\Delta(n,\mu)\simeq n^{-1}C[\varepsilon(n,\mu)]+O(\mu^{-1}).
\end{equation}

\subsection{Technical errors}

In WTB the crucial technical error is clearly the treatment of the
\textquotedblleft infidelity\textquotedblright\ parameter $\varepsilon
_{\text{TP}}$ which appears in Eq.~(\ref{errorTT}) and is defined as the
infidelity between the outputs of the protocol $\rho_{\mathbf{ab}}^{n}$\ and
the simulated protocol $\rho_{\mathbf{ab}}^{\mu,n}$ [in WTB denoted these as
$\zeta_{AB}^{n}$ and $\zeta_{AB}^{\prime}(n,\mu)$]. More precisely, this
is~\cite[Eq.~(177)]{WildeFollowup}
\begin{equation}
\varepsilon_{\text{TP}}(n,\mu):=1-F(\rho_{\mathbf{ab}}^{n},\rho_{\mathbf{ab}%
}^{\mu,n}), \label{pb1}%
\end{equation}
where $F$ is the quantum fidelity. WTB argues that~\cite{WildeFollowup}%
\begin{equation}%
\begin{array}
[c]{c}%
\text{\textquotedblleft\textit{continuous variable\ teleportation}}\\
\text{\textit{induces a perfect quantum channel}}\\
\text{\textit{when infinite energy is available}\textquotedblright}%
\end{array}
\label{WTBstates}%
\end{equation}
which is the only reason why WTB states~\cite[Eq.~(178)]{WildeFollowup}%
\begin{equation}
\underset{\mu\rightarrow\infty}{\lim\sup}~\varepsilon_{\text{TP}}%
(n,\mu)=0,~~\text{for any }n\text{.} \label{pb2}%
\end{equation}

The first error is a basic misinterpretation of the convergence properties of
the BK\ protocol. In fact, the statement~(\ref{WTBstates}) clearly means that
the BK teleportation channel $\mathcal{I}^{\mu}$ generated by performing the
protocol over a finite energy TMSV state $\Phi^{\mu}$ would reproduce the
identity channel $\mathcal{I}$ (\textquotedblleft perfect quantum
channel\textquotedblright) when $\mu\rightarrow\infty$. We know that this is
not true. In fact, as we have already shown in Eq.~(\ref{nonUNI}), we have%
\begin{equation}
\lim_{\mu\rightarrow\infty}\left\Vert \mathcal{I}^{\mu}-\mathcal{I}\right\Vert
_{\diamond}=2~.
\end{equation}
In other words, the BK channel does not converge to the identity channel, as
explained in detail in Sec.~\ref{nonUNIsec}\ and already pointed out in
Ref.~\cite{TQC}. Unfortunately, this has catastrophic consequences for the
statement in Eq.~(\ref{pb2}) and all the WTB claims.

To make it simple, consider the single use ($n=1$) of a trivial adaptive
protocol ($\Lambda_{1}=\mathcal{I}$) performed over channel $\mathcal{C}$ and
its teleportation simulation $\mathcal{C}^{\mu}=\mathcal{C}\circ
\mathcal{I}^{\mu}$. From Eqs.~(\ref{ada1}) and~(\ref{ada22}), we have the two
output states
\begin{equation}
\rho_{\mathbf{ab}}^{1}=\mathcal{C}(\rho_{\mathbf{ab}}^{0}),~\rho_{\mathbf{ab}%
}^{\mu,1}=\mathcal{C}^{\mu}(\rho_{\mathbf{ab}}^{0}),
\end{equation}
where the channels are meant to be applied to the input system $a_{1}$, i.e.,
$\mathcal{C}=\mathcal{I}_{\mathbf{a}}\otimes\mathcal{C}_{a_{1}}\otimes
\mathcal{I}_{\mathbf{b}}$ and $\mathcal{C}^{\mu}=\mathcal{I}_{\mathbf{a}%
}\otimes\mathcal{C}_{a_{1}}^{\mu}\otimes\mathcal{I}_{\mathbf{b}}$. The
infidelity is given by%
\begin{align}
\varepsilon_{\text{TP}}(1,\mu)  &  =1-F(\rho_{\mathbf{ab}}^{1},\rho
_{\mathbf{ab}}^{\mu,1})\\
&  =1-F[\mathcal{C}_{a_{1}}(\rho_{\mathbf{ab}}^{0}),\mathcal{C}_{a_{1}}^{\mu
}(\rho_{\mathbf{ab}}^{0})]\\
&  \geq1-F[\rho_{\mathbf{ab}}^{0},\mathcal{I}_{a_{1}}^{\mu}(\rho_{\mathbf{ab}%
}^{0})], \label{eqMMM}%
\end{align}
where we have exploited the monotonicity of the fidelity under CPTP maps,
considering $\mathcal{C}=\mathcal{C}\circ\mathcal{I}$ and $\mathcal{C}^{\mu
}=\mathcal{C}\circ\mathcal{I}^{\mu}$.

The proof idea in WTB was the exploitation of the (wrong) uniform limit
$\mathcal{I}_{a_{1}}^{\mu}\overset{\mu}{\rightarrow}\mathcal{I}_{a_{1}}$ [see
the statement in~(\ref{WTBstates})], so that one could write $\lim_{\mu}%
F[\rho_{\mathbf{ab}}^{0},\mathcal{I}_{a_{1}}^{\mu}(\rho_{\mathbf{ab}}^{0})]=1$
in Eq.~(\ref{eqMMM}). Instead, assume that Alice is sending part of a TMSV
state $\Phi^{\tilde{\mu}}$ with energy $\tilde{\mu}$. This means that we may
decompose $\rho_{\mathbf{ab}}^{0}=\rho_{\mathbf{a}}^{0}\otimes\Phi_{aa_{1}%
}^{\tilde{\mu}}\otimes\rho_{\mathbf{b}}^{0}$, and write%
\begin{equation}
\varepsilon_{\text{TP}}(1,\mu)\geq1-F[\Phi_{aa_{1}}^{\tilde{\mu}}%
,\mathcal{I}_{a}\otimes\mathcal{I}_{a_{1}}^{\mu}(\Phi_{aa_{1}}^{\tilde{\mu}%
})], \label{firstCC}%
\end{equation}
where we use the multiplicativity of the fidelity under tensor products.
Taking the $\lim\sup_{\mu}$ of $\varepsilon_{\text{TP}}(1,\mu)$ means to
include all the possible input states. Because the input alphabet is unbounded
(as it should be when we consider unconstrained quantum and private
capacities), this means that the alphabet also includes the limit of
asymptotic states, such as $\Phi_{aa_{1}}^{\tilde{\mu}}$ for large $\tilde
{\mu}$.

Also note that the generic limit in Eq.~(\ref{pb2}) does not imply any
specific order of the limits between the simulation energy $\mu$ and the input
energy $\tilde{\mu}$ of the alphabet. For an unbounded alphabet, we can
equivalently interpret%
\begin{equation}
\underset{\mu}{\lim\sup}=\lim_{\mu}\lim_{\tilde{\mu}}\text{~~OR~}%
~\underset{\mu}{\lim\sup}=\lim_{\tilde{\mu}}\lim_{\mu}~. \label{WTBambiguity}%
\end{equation}
Therefore, if we apply the first case to Eq.~(\ref{firstCC}) we find%
\begin{align}
&  \underset{\mu\rightarrow\infty}{\lim\sup~}\varepsilon_{\text{TP}}%
(1,\mu)\nonumber\\
&  \geq1-\lim_{\mu\rightarrow\infty}\lim_{\tilde{\mu}\rightarrow\infty}%
F[\Phi_{aa_{1}}^{\tilde{\mu}},\mathcal{I}_{a}\otimes\mathcal{I}_{a_{1}}^{\mu
}(\Phi_{aa_{1}}^{\tilde{\mu}})]\nonumber\\
&  \geq1~,
\end{align}
because, as we know, $F[\Phi_{aa_{1}}^{\tilde{\mu}},\mathcal{I}_{a}%
\otimes\mathcal{I}_{a_{1}}^{\mu}(\Phi_{aa_{1}}^{\tilde{\mu}})]\simeq
O(\tilde{\mu}^{-1})$ at any fixed $\mu$. This result disproves the claim in
Eq.~(\ref{pb2}) already for the trivial case of $n=1$.

\begin{remark}
It is important to remark that the ambiguity in Eq.~(\ref{WTBambiguity}) is
not addressed, discussed or noted in any part of WTB, where the convergence
problems of the BK protocol are just completely ignored. In WTB there is no
discussion related to uniform convergence [associated with the first order of
the limits in Eq.~(\ref{WTBambiguity})] or strong convergence [associated with
the second order of the limits in Eq.~(\ref{WTBambiguity})]. Also note that
the additional arguments presented in the various arXiv versions of
Ref.~\cite[v1-v5]{WildePLAG} can be seen as an erratum de facto of WTB, rather
than a justification of its proofs as claimed by the author.
\end{remark}

As the \textquotedblleft proof\textquotedblright\ has been carried out in WTB,
one must conclude that
\begin{equation}
\underset{\mu\rightarrow\infty}{\lim\sup}~\varepsilon_{\text{TP}}%
(n,\mu)=1,~~\text{for any }n\text{,} \label{contro}%
\end{equation}
which is exactly the opposite of the claim in Eq.~(\ref{pb2}). In fact, one
can extend the previous reasoning to any $n$ and any adaptive protocol (which
is the content of the next section). The result in Eq.~(\ref{contro}) implies
$\lim\sup_{\mu}~\varepsilon(n,\mu)=1$ for the overall error in
Eq.~(\ref{errorTT}). Unfortunately, this leads to $C[\varepsilon
(n,\mu)]\rightarrow\infty$ in Eq.~(\ref{eqrr}) and, therefore, all the bounds
claimed by WTB in Eq.~(\ref{followPLOB}) are divergent, i.e.,
\begin{equation}
K(\mathcal{C})\leq\Phi(\mathcal{C})+\infty~.
\end{equation}

\subsection{Filling the technical gaps}

It is important to note that, in WTB, the simulation error on the output state
$\varepsilon_{\text{TP}}(n,\mu)$ is completely disconnected from the error on
the channel simulation which affects each transmission. In other words, there
are no rigorous relations such as those given in Eqs.~(\ref{OOO1}),
(\ref{OOO2}) or (\ref{OOO3}) for the various forms of convergence. The reason
is because in WTB there is no peeling argument~\cite{PLOB,TQC} which
simplifies the adaptive protocol and relates the output error $\left\Vert
\rho_{\mathbf{ab}}^{\mu,n}-\rho_{\mathbf{ab}}^{n}\right\Vert $ to the channel
error $\mathcal{C}^{\mu}\neq\mathcal{C}$. As a result, the WTB claims not only
are not proven (due to the divergences) but they \textit{do not even apply} to
adaptive protocols. Here we apply the peeling argument to correctly write
$\varepsilon_{\text{TP}}(n,\mu)$ in the presence of an adaptive protocol. This
extends the considerations already made in Ref.~\cite{TQC} for the
bounded-uniform convergence to the other forms of convergence (strong and uniform).

Using the Fuchs-van der Graaf relations of Eq.~(\ref{FuchsGraaf}), we may
write
\begin{equation}
\varepsilon_{\text{TP}}(n,\mu)\leq\frac{\left\Vert \rho_{\mathbf{ab}}^{n}%
-\rho_{\mathbf{ab}}^{\mu,n}\right\Vert }{2}~.
\end{equation}
Following PLOB and Ref.~\cite{TQC}, we may consider the energy-constrained
diamond distance and perform the peeling procedure in Eqs.~(\ref{peel11}%
)-(\ref{casen2}) which leads to the result in Eq.~(\ref{OOO1}). Therefore, we
may write%
\begin{equation}
\varepsilon_{\text{TP}}(n,\mu)\leq n\delta(\mu,N)/2,
\end{equation}
where $\delta(\mu,N):=\left\Vert \mathcal{C}^{\mu}-\mathcal{C}\right\Vert
_{\diamond N}$. Once we have the control on the error, we may take the limit
for large $\mu$. For any number of channel uses $n$ and \textit{finite} energy
constraint $N$, we may safely write%
\begin{equation}
\underset{\mu\rightarrow\infty}{\lim\sup}~\varepsilon_{\text{TP}}(n,\mu
|N)\leq\lim_{\mu\rightarrow\infty}~n\delta(\mu,N)/2=0~, \label{newf}%
\end{equation}
proving Eq.~(\ref{pb2}) and the corresponding WTB claims.

More precisely, starting from Eq.~(\ref{newf}), we may write the following
upper bound for the \textit{energy-constrained} key capacity~\cite{TQC}
\begin{equation}
K(\mathcal{C}|N)\leq\Phi(\mathcal{C})+\Delta(n,\mu|N)~, \label{UBkk}%
\end{equation}
where $\Delta(n,\mu|N)$ is computed assuming the energy constraint. For large
$\mu$, we may now write
\begin{equation}
\Delta(n,\mu|N)\rightarrow\sqrt{\frac{V(\mathcal{C})}{n(1-\varepsilon)}}%
+\frac{C(\varepsilon)}{n}.
\end{equation}
Because $\lim_{\mu}\Delta(n,\mu|N)$ does not depend on $N$, we can extend the
inequality in Eq.~(\ref{UBkk}) to the supremum $K(\mathcal{C}):=\sup
_{N}K(\mathcal{C}|N)$, so that
\begin{equation}
K(\mathcal{C})\leq\Phi(\mathcal{C})+\lim_{\mu\rightarrow\infty}\Delta
(n,\mu|N),
\end{equation}
proving the strong converse bound claimed in Eq.~(\ref{strongconv}).

Additional proofs can be made assuming the other types of convergence (strong
and uniform). These are simple variants of the previous one. First of all,
because we consider phase-insensitive canonical forms, we now know that the
teleportation simulation converges uniformly, i.e., $\left\Vert \mathcal{C}%
^{\mu}-\mathcal{C}\right\Vert _{\diamond}\overset{\mu\rightarrow\infty
}{\rightarrow}0$. This means that we may directly consider $N=\infty$ in the
previous proof, so that we can delete the conditioning from the energy
constraint $N$ and the last step (supremum in $N)$ is not needed. Another
approach is considering the strong convergence of the teleportation
simulation. After the peeling procedure, this means that we may write
Eq.~(\ref{OOO3}), so that
\begin{equation}
\varepsilon_{\text{TP}}(n,\mu)\leq\frac{n}{2}\sup_{\rho_{\mathbf{ab}}}%
\Vert\mathcal{C}(\rho_{\mathbf{ab}})-\mathcal{C}^{\mu}(\rho_{\mathbf{ab}%
})\Vert\overset{\mu\rightarrow\infty}{\rightarrow}0.
\end{equation}

\begin{remark}
It is easy to check that none of these techniques have been explicitly or even
implicitly discussed in WTB, contrary to what claimed in the various arXiv
versions of Ref.~\cite[v1-v5]{WildePLAG}.
\end{remark}

\section{Conclusions}

In this work we have discussed the Braunstein-Kimble teleportation protocol
for bosonic systems and its application to the simulation of bosonic channels.
We have considered the various forms (topologies) of convergence of this
protocol to the identity channel, which are still the subject of basic
misunderstandings for some authors. As a completely new result, we have shown
that the teleportation simulation of an arbitrary single-mode Gaussian channel
(not necessarily in canonical form) uniformly converges to the channel in the
limit of infinite energy, as long as the channel has a full rank noise matrix.

We have then discussed the various forms of convergence in the context of
adaptive protocols, following the ideas established in PLOB~\cite{PLOB}. In
this scenario, it is essential to provide a peeling procedure which relates
the simulation error on the final output state to the simulation error
affecting the individual channel transmissions. As an application, we exploit
this peeling argument and the various convergence topologies to completely
prove the claims presented in WTB in relation to private communication over
bosonic Gaussian channels. This treatment extends the first rigorous proof
given in Ref.~\cite{TQC} and specifically based on the bounded-uniform
convergence (energy-constrained diamond distance).

\bigskip

\textbf{Acknowledgments.}~ This research has been funded by the EPSRC via the
`UK Quantum Communications HUB' (Grant no. EP/M013472/1) and the Innovation
Fund Denmark (Qubiz project). The authors would like to thank discussions with
C. Ottaviani, T. P. W. Cope, G. Spedalieri, and L. Banchi.

\bigskip

\textbf{Author contribution statement.}~The authors contributed
equally to the manuscript.

\appendix

\section{Proof of Lemma~\ref{LemmaMAIN}\label{APPproof}}

Consider the canonical forms $\mathcal{C}$ with $\tau:=\det\mathbf{T}\neq1$
and $\text{rank}(\mathbf{N})=2$. These correspond to $A_{2}$, $C($Att$)$,
$C($Amp$)$, and $D$. Given $\mathcal{C}$, consider the variant%
\begin{equation}
\mathcal{C}^{\mu}:=\mathcal{C}\circ\mathcal{U}_{A}\circ\mathcal{I}^{\mu}%
\circ\mathcal{U}_{A}^{-1}, \label{fromEQ}%
\end{equation}
where $\mathcal{U}_{A}$ is a canonical Gaussian unitary with associated
symplectic matrix $\mathbf{S}_{A}$, and $\mathcal{I}^{\mu}$ is the
BK\ teleportation channel, which is locally (point-wise) equivalent to an
additive-noise Gaussian channel ($B_{2}$ form) with added noise
\begin{equation}
\xi=2[\mu-\sqrt{\mu^{2}-1}]. \label{defcsi}%
\end{equation}
Note that we may use the Bloch-Messiah decomposition~\cite{SamMessiah}
\begin{equation}
\mathbf{S}_{A}=\mathbf{O}_{1}\mathbf{S}_{q}\mathbf{O}_{2}, \label{Messiah}%
\end{equation}
where $\mathbf{O}$'s are symplectic orthogonal matrices, while $\mathbf{S}%
_{q}=\mathrm{diag}(r,r^{-1})$ for $r>0$ is a squeezing matrix~\cite{Euler}.
Here we show that $\mathcal{C}$ and $\mathcal{C}^{\mu}$ have the same unitary
dilation with different environmental states $\rho_{e}$ and $\rho_{e}^{\mu}$,
whose fidelity $F(\rho_{e}^{\mu},\rho_{e})\overset{\mu\rightarrow\infty
}{\rightarrow}1$. Let us start with the form $C$.

\subsection{Lossy channel $C($Att$)$ and amplifier $C($Amp$)$}

Consider the canonical $C$ form $\mathcal{C}(\tau>0,2,\bar{n})$ representing
either a thermal-loss channel ($0<\tau<1$) or a noisy quantum amplifier
($\tau>1$). Their action on the input covariance matrix (CM) $\mathbf{V}$ is
given by%
\begin{equation}
\mathbf{V}\rightarrow\tau\mathbf{V}+|1-\tau|\omega\mathbf{I}~,
\end{equation}
where $\omega:=2\bar{n}+1\geq1$. From Eq.~(\ref{fromEQ}), we may write
\begin{align}
\mathbf{V}  &  \rightarrow\tau(\mathbf{V}+\xi\mathbf{S}_{A}\mathbf{S}_{A}%
^{T})+|1-\tau|\omega\mathbf{I}\nonumber\\
&  =\tau\mathbf{V}+|1-\tau|\mathbf{\tilde{W}}, \label{Cov1}%
\end{align}
where we have set
\begin{equation}
\mathbf{\tilde{W}}:=\omega\mathbf{I}+\gamma\mathbf{S}_{A}\mathbf{S}_{A}%
^{T},~~\gamma:=\frac{\xi\tau}{|1-\tau|}\geq0. \label{envMM}%
\end{equation}
According to Eqs.~(\ref{Cov1}) and~(\ref{envMM}), we may represent
$\mathcal{C}^{\mu}(\tau>0,2,\bar{n})$ with the same two-mode symplectic matrix
$\mathbf{M}(C)$ of the original $C$ form, but replacing the thermal state
$\rho_{e}(\bar{n})$ with a zero-mean Gaussian state $\rho_{e}^{\mu}$ whose CM
can be written as $\mathbf{\tilde{W}}$. To check this is indeed the case, we
need to verify that $\mathbf{\tilde{W}}$ is a bona fide CM ~\cite{bonafide}.
It is certainly positive definite, so we just need to check that its
symplectic eigenvalue is greater than $1$. Note that we may apply the
orthogonal symplectic $\mathbf{O}_{1}$ so that
\begin{equation}
\mathbf{W:}=\mathbf{O}_{1}^{T}\mathbf{\tilde{W}O}_{1}=\omega\mathbf{I}%
+\gamma\mathbf{S}_{q}^{2}~.
\end{equation}
The symplectic eigenvalue is equal to
\begin{align}
\nu &  =\sqrt{\det\mathbf{W}}=\sqrt{\omega^{2}+\gamma^{2}+\gamma\omega\left(
r^{2}+1/r^{2}\right)  }\nonumber\\
&  \geq\omega+\gamma\geq1~.
\end{align}
Finally we compute the fidelity between the environmental states, finding
\begin{gather}
F(\rho_{e}^{\mu},\rho_{e})=\sqrt{2r}\left[  \sqrt{\left(  \gamma r^{2}%
\omega+\omega^{2}+1\right)  \left(  \gamma\omega+r^{2}\left(  \omega
^{2}+1\right)  \right)  }\right. \nonumber\\
-\left.  \sqrt{\left(  \omega^{2}-1\right)  \left(  \gamma\omega+\gamma
r^{4}\omega+r^{2}\left(  \gamma^{2}+\omega^{2}-1\right)  \right)  }\right]
^{-1/2}, \label{FCAMP}%
\end{gather}
which goes to $1$ for $\mu\rightarrow\infty$ (so that $\xi\rightarrow0$ and
$\gamma\rightarrow0$). This is true for any finite value of the squeezing
$r>0$ and the thermal variance $\omega$.

\subsection{Conjugate of the amplifier $D$}

Let us consider the $D$ form $\mathcal{C}(\tau<0,2,\bar{n})$ which transforms
the input as follows%
\begin{equation}
\mathbf{V}\rightarrow-\tau\mathbf{ZVZ}+(1-\tau)\omega\mathbf{I}.
\end{equation}
Then, the action of $\mathcal{C}^{\mu}(\tau<0,2,\bar{n})$ can be written as
\begin{align}
\mathbf{V}  &  \rightarrow-\tau\mathbf{Z}(\mathbf{V}+\xi\mathbf{S}%
_{A}\mathbf{S}_{A}^{T})\mathbf{Z}+(1-\tau)\omega\mathbf{I}\nonumber\\
&  =-\tau\mathbf{Z}\mathbf{V}\mathbf{Z}+(1-\tau)\left(  \omega\mathbf{I}%
-\kappa\mathbf{Z}\mathbf{S}_{A}\mathbf{S}_{A}^{T}\mathbf{Z}\right) \nonumber\\
&  =-\tau\mathbf{Z}\mathbf{V}\mathbf{Z}+(1-\tau)\mathbf{\tilde{W}}
\label{Cov2}%
\end{align}
where $\kappa:=\xi\tau/(1-\tau)\leq0$. Using the Bloch-Messiah decomposition
of Eq.~(\ref{Messiah}) and $\mathbf{ZS}_{q}^{2}\mathbf{Z}=\mathbf{S}_{q}^{2}$,
we may write
\begin{align}
\mathbf{\tilde{W}}  &  =\omega\mathbf{I}-\kappa\mathbf{ZO}_{1}\mathbf{S}%
_{q}^{2}\mathbf{O}_{1}^{T}\mathbf{Z}\nonumber\\
&  =\omega\mathbf{I}-\kappa(\mathbf{ZO}_{1}\mathbf{Z)S}_{q}^{2}(\mathbf{ZO}%
_{1}^{T}\mathbf{Z).} \label{Wtilde}%
\end{align}
Thus, we may represent $\mathcal{C}^{\mu}(\tau<0,2,\bar{n})$ with the same
two-mode symplectic matrix $\mathbf{M}(D)$ as the original $D$ form, but
replacing the thermal state $\rho_{e}(\bar{n})$ with a zero-mean Gaussian
state $\rho_{e}^{\mu}$ whose CM can be written as $\mathbf{\tilde{W}}$ in
Eq.~(\ref{Wtilde}). To check this is indeed the case, we need to verify that
$\mathbf{\tilde{W}}$ is a bona fide CM~\cite{bonafide}. First notice that the
matrix $\boldsymbol{\Sigma}:=\mathbf{ZO}_{1}\mathbf{Z}$ is orthogonal and
symplectic. We may therefore apply the symplectic $\boldsymbol{\Sigma}^{T}$
and write
\[
\mathbf{W}=\boldsymbol{\Sigma}^{T}\mathbf{\tilde{W}}\boldsymbol{\Sigma}%
=\omega\mathbf{I}-\kappa\mathbf{S}_{q}^{2}~.
\]
Because $\kappa\leq0$, this is positive definite and it has symplectic
eigenvalue
\begin{align}
\nu &  =\sqrt{\omega^{2}+\kappa^{2}-\omega\kappa(r^{2}+1/r^{2})}\nonumber\\
&  \geq\omega-\kappa\geq1~.
\end{align}
Finally we compute the fidelity between the environmental states, finding%
\begin{gather}
F(\rho_{e}^{\mu},\rho_{e})=\sqrt{2r}\left[  \sqrt{\left(  -\kappa r^{2}%
\omega+\omega^{2}+1\right)  \left(  -\kappa\omega+r^{2}\left(  \omega
^{2}+1\right)  \right)  }\right. \nonumber\\
\left.  -\sqrt{\left(  1-\omega^{2}\right)  \left(  \kappa\omega+\kappa
r^{4}\omega-r^{2}\left(  \kappa^{2}+\omega^{2}-1\right)  \right)  }\right]
^{-1/2},
\end{gather}
which goes to $1$ for large $\mu$ (so that $\xi\rightarrow0$ and
$\kappa\rightarrow0$). This is true for any finite value of the squeezing
$r>0$ and the thermal variance $\omega$.

\subsection{Canonical form $A_{2}$}

The $A_{2}$ form\ $\mathcal{C}(0,1,\bar{n})$ transforms the input CM as
\begin{equation}
\mathbf{V}\rightarrow\boldsymbol{\Pi}\mathbf{V}\boldsymbol{\Pi}+\omega
\mathbf{I~,}%
\end{equation}
where
\begin{equation}
\boldsymbol{\Pi}:=\frac{\mathbf{I}+\mathbf{Z}}{2}=\mathrm{diag}(1,0).
\end{equation}
The action of the variant $\mathcal{C}^{\mu}(0,1,\bar{n})$ is given by
\begin{equation}
\mathbf{V}\rightarrow\boldsymbol{\Pi(}\mathbf{V+\xi\mathbf{S}_{A}%
\mathbf{S}_{A}^{T})}\boldsymbol{\Pi}+\omega\mathbf{I}=\boldsymbol{\Pi
}\mathbf{V}\boldsymbol{\Pi}+\mathbf{\tilde{W}}~, \label{Cov3}%
\end{equation}
where
\begin{equation}
\mathbf{\tilde{W}}:=\omega\mathbf{I+}\xi\boldsymbol{\Pi}\mathbf{\mathbf{S}%
_{A}\mathbf{S}_{A}^{T}}\boldsymbol{\Pi}. \label{A2GA}%
\end{equation}
Thus, we may represent $\mathcal{C}^{\mu}(0,1,\bar{n})$ with the same two-mode
symplectic matrix $\mathbf{M}(A_{2})$ of the original $A_{2}$ form, but
replacing the thermal state $\rho_{e}(\bar{n})$ with a zero-mean Gaussian
state $\rho_{e}^{\mu}$ whose CM can be written as $\mathbf{\tilde{W}}$ in
Eq.~(\ref{A2GA}). To check this is indeed the case, we need to verify that
$\mathbf{\tilde{W}}$ is a bona fide CM~\cite{bonafide}. $\mathbf{\tilde{W}}$
is clearly positive definite. To derive its symplectic eigenvalue, let us set
\begin{equation}
\mathbf{S}_{A}=\left(
\begin{array}
[c]{cc}%
a & c\\
d & b
\end{array}
\right)  ,
\end{equation}
where the real entries must satisfy $\det\mathbf{S}_{A}=ab-cd=1$. Then we get
\begin{equation}
\mathbf{\tilde{W}}=\left(
\begin{array}
[c]{cc}%
\xi(a^{2}+c^{2})+\omega & 0\\
0 & \omega
\end{array}
\right)  ~,
\end{equation}
with symplectic eigenvalue
\begin{equation}
\nu=\sqrt{[\xi(a^{2}+c^{2})+\omega]\omega}\geq\omega\geq1~.
\end{equation}
Finally we compute the fidelity between the environmental states, yielding%
\begin{gather}
F(\rho_{e}^{\mu},\rho_{e})=\sqrt{2}\left[  \sqrt{\left(  \omega^{2}+1\right)
\left(  \xi\omega\left(  a^{2}+c^{2}\right)  +\omega^{2}+1\right)  }\right.
\nonumber\\
\left.  -\sqrt{\left(  \omega^{2}-1\right)  \left(  \xi\omega\left(
a^{2}+c^{2}\right)  +\omega^{2}-1\right)  }\right]  ^{-1/2},
\end{gather}
which clearly goes to $1$ for large $\mu$ (i.e., for $\xi\rightarrow0$). This
is true for any finite value of the real parameters $a$ and $c$, and the
thermal variance $\omega$.


\section{Asymptotic results for the $B_{2}$ form\label{AsyAPP}}

Consider the $B_{2}$ form $\mathcal{C}[1,2,\xi^{\prime}]$ with added noise
$\xi^{\prime}$. This can be expressed as an asymptotic $C$ form $\mathcal{C}%
(0<\tau<1,2,\bar{n})$ with $\tau\rightarrow1$ and thermal variance%
\begin{equation}
\omega=\xi^{\prime}/(1-\tau). \label{xc}%
\end{equation}
The channel $\mathcal{C}[1,2,\xi^{\prime}]$ and its simulation $\mathcal{C}%
^{\mu}[1,2,\xi^{\prime}]$ [according to Eq.~(\ref{fromEQ})] have the same
(asymptotic)\ unitary dilation but different environmental states $\rho_{e}$
and $\rho_{e}^{\mu}$. These are the states associated with $\mathcal{C}%
(0<\tau<1,2,\bar{n}_{\xi^{\prime},\tau})$ and $\mathcal{C}^{\mu}%
(0<\tau<1,2,\bar{n}_{\xi^{\prime},\tau})$ where $\bar{n}_{\xi^{\prime},\tau
}:=[\xi^{\prime}(1-\tau)^{-1}-1]/2$. Using Eq.~(\ref{xc}) in Eq.~(\ref{FCAMP})
and taking the limit for $\tau\rightarrow1$, we may write%
\begin{equation}
F(\rho_{e}^{\mu},\rho_{e})=2\sqrt{\frac{r\xi^{\prime}\sqrt{\xi\xi^{\prime
}+r^{4}\xi\xi^{\prime}+r^{2}(\xi^{2}+\xi^{\prime2})}}{2\xi\xi^{\prime}%
(1+r^{4})+r^{2}(\xi^{2}+4\xi^{\prime2})}}+O(\tau-1),
\end{equation}
where $\xi$ is defined in Eq.~(\ref{defcsi}) and $r$ is a squeezing parameter
associated with the input canonical unitary $\mathcal{U}_{A}$. Then, the limit
in $\xi\rightarrow0$ (i.e., $\mu\rightarrow\infty$) provides%
\begin{equation}
F(\rho_{e}^{\mu},\rho_{e})=1+O(\xi)+O(\tau-1)~.
\end{equation}
Similarly, we may write the expansion%
\begin{equation}
2\sqrt{1-F(\rho_{e}^{\mu},\rho_{e})^{2}}=O(\xi)+O(\tau-1).
\end{equation}


\begin{thebibliography}{99}
\bibitem {Weyl}Notice that our choice of the Weyl operator is different from
$\hat{D}(\boldsymbol{\xi}):=\exp(i\mathbf{\hat{x}}^{T}\mathbf{\Omega
}\boldsymbol{\xi})$, whose definition directly comes from the standard
displacement operator of quantum optics. The choice of
Eq.~(\ref{Weyl operator}) avoids the presence of $\mathbf{\Omega}$ in the
characteristic functions.
\bibitem {notation}Note that here we are considering different quantum
shot-noise units with respect to Refs.~\cite{HolevoVittorio,Caruso}. In our
case we have quadratures of the form $\hat{q}:=(\hat{a}+\hat{a}^{\dagger})$
and $\hat{p}:=-i(\hat{a}-\hat{a}^{\dagger})$, so that $[\hat{a},\hat
{a}^{\dagger}]=1$\ implies the commutation relation $[\hat{q},\hat{p}]=2i$. As
a consequence, the noise variance of the vacuum state is equal to $1$. We
follow the notation of Ref.~\cite{RMP}.
\bibitem {NoteSympl}In fact, since $\mathbf{M}$ is symplectic we have
$\det\mathbf{m}_{1}+\det\mathbf{m}_{2}=1$. Then, $\det\mathbf{N}_{c}=(2\bar
{n}+1)^{2}(\det\mathbf{m}_{2})^{2}=(2\bar{n}+1)^{2}(\det\mathbf{m}_{1}%
-1)^{2}=(2\bar{n}+1)^{2}(\det\mathbf{T}_{c}-1)^{2}\geq(\det\mathbf{T}%
_{c}-1)^{2}$.
\bibitem {Euler}In general, any symplectic transformation $\mathbf{S}%
\in\mathcal{S}p(2n,\mathbb{R})$ admits the decomposition $\mathbf{S}%
=\mathbf{K~}%
{\textstyle\bigoplus\nolimits_{k=1}^{n}}
\mathbf{S}_{q}(r_{k})~\mathbf{L}$, where $\mathbf{S}_{q}(r_{k}):=\mathrm{diag}%
(r_{k},r_{k}^{-1})$ are single-mode squeezers, while $\mathbf{K}$ and
$\mathbf{L}$ are passive transformations belonging to the compact subgroup
$\mathcal{K}(2n)=\mathcal{S}p(2n,\mathbb{R})\cap\mathcal{SO}(2n)$. In
particular, for $n=1$, we simply have $\mathcal{K}(2)=\mathcal{SO}(2)$, i.e.,
all the proper rotations in $\mathbb{R}^{2}$ are canonical (simply because
$\mathbf{S}\in\mathcal{S}p(2,\mathbb{R})\Leftrightarrow\det\mathbf{S}=+1$ for
$n=1$).
\end{thebibliography}
\end{document}